\documentclass[12pt,a4paper]{article}
\pdfoutput=1

\usepackage{jheppub}
\usepackage{dsfont}
\usepackage{amssymb,amsmath}
\usepackage{epsfig}
\usepackage{epstopdf}
\usepackage{latexsym}
\usepackage{graphicx}
\usepackage{subfig}
\usepackage{booktabs}
\usepackage{bbm}
\usepackage{color}

\makeatletter
\def\@fpheader{\relax}
\makeatother

\def\be{\begin{equation}}
\def\ee{\end{equation}}

\def\S{{\cal S}}
\def\A{{\cal A}}
\def\R{{\cal R}}

\def\O{{\cal O}}

\def\S{{\cal S}}

\def\R{{\mathbb R}}

\preprint{YITP-16-146}

\title{Fundamental Flavours, Fields and Fixed Points: A Brief Account}
\author{Arnab Kundu$^{a}$, Nilay Kundu$^c$}
\affiliation{$^{a}$Theory Division, Saha Institute of Nuclear Physics (HBNI), 1/AF Bidhannagar, Kolkata 700064, India.}
\affiliation{$^c$Center for Gravitational Physics, Yukawa Institute for Theoretical Physics (YITP), Kyoto University, Kyoto 606-8502, Japan.}
\emailAdd{arnab.kundu [at] saha.ac.in, nilay.tifr[at]gmail.com}

\abstract{In this article we report on a preliminary study, {\it via} Holography, of infrared fixed points in a putative strongly coupled SU$(N_c)$ gauge theory, with $N_f$ fundamental matter, in the presence of additional fields in the fundamental sector, {\it e.g.}~density or a magnetic field. In an inherently {\it effective} or a {\it bottom up} approach, we work with a simple system: Einstein-gravity with a negative cosmological constant, coupled to a Dirac-Born-Infeld (DBI) matter. We obtain a class of exact solutions, dual to candidate grounds states in the infrared (IR), with a scaling ansatz for various fields. These solutions are of two kinds: AdS$_m \times {\mathbb R}^n$--type, which has appeared in the literature before; and AdS$_m \times$EAdS$_n$--type, where $m$ and $n$ are suitable integers. Both these classes of solutions are {\it non-perturbative} in back-reaction. The AdS$_m \times$EAdS$_n$--type contains examples of Bianchi type-V solutions. We also construct explicit numerical flows from an AdS$_5$ ultraviolet to both an AdS$_2$ and an AdS$_3$ IR.}

\begin{document}

\maketitle
\flushbottom

\section{Introduction \& Conclusions}

The study of Renormalization Group (RG) fixed points within the framework of quantum field theory (QFT) has been remarkably rich, fruitful, alluring and incisive to universality across various realms of physics. Among these, one of the most sought after are theories that are quantum chromodynamics (QCD)-like, in which matter fields in adjoint (gluons) and fundamental (quarks) representation of an SU$(N_c)$-gauge group constitute the degrees of freedom. An early study in QCD-like theories that reveal a vanishing beta-function, in presence of both adjoint and fundamental matter fields, is the so-called Caswell-Banks-Zaks fixed point\cite{Caswell:1974gg, Banks:1981nn}.

The class of asymptotically free QCD-like theories has been under much scrutiny as a function of the number of flavours, or more precisely, as a function of the ratio of number of flavours, $N_f$, and the number of colours, $N_c$. It is known that depending on $N_f/N_c$, there is a {\it conformal window}: a region in the theory parameter space. Within the conformal window, the corresponding infrared physics is governed by a non-trivial fixed point, which the RG-flow leads to. Such behaviour generalizes to supersymmetric theories, as well. Perhaps it is a good place to mention that, if not all\footnote{For example, in supersymmetric theories, certain perturbative results are exact.}, much of these studies are perturbative in some loop expansion.

Given the already {\it flavoured}-richness, there are, however, rather outstanding questions that confront current theoretical tools, and understanding. One of those is the understanding of the ground state of a QCD-like theory with non-vanishing density (or, chemical potential) at strong coupling. A perturbative approach is not useful; lattice techniques are, at best, limited at non-vanishing density, due to the so-called ``sign-probem". Though it is possible to construct supersymmetric theories with all desired ingredients, that perhaps remains tractable to exact and analytical results, it seems to be a still less-explored avenue.

We will, instead, take a different route: We want to view the Gauge-String duality, or the AdS/CFT correspondence\cite{Maldacena:1997re} as a framework of studying quantum field theories, at strong coupling. In \cite{Karch:2002sh}, fundamental matter field was introduced by virtue of explicitly introducing a D-brane probe in a background geometry. This brane was introduced in the so-called {\it probe limit}, in which $N_f \ll N_c$, and the geometry does not receive any correction due to the brane source. While this limit has, since then, been explored in details (see {\it e.g.}~\cite{Kobayashi:2006sb, Karch:2007br}), relatively less is understood away from the probe limit. On the other hand, physically, the flavour back-reaction is rather interesting, specially in view of the possibility of exotic states such as {\it colour superconductivity} at high density\cite{Alford:2007xm}. We note that a large and extensive literature on back-reaction by fundamental flavours already exists, which we will not attempt to enlist here. For our current purpose, we will specifically cherrypick a few observations of \cite{Faedo:2014ana, Faedo:2015ula, Faedo:2015urf, Faedo:2016jbd}. The infrared (IR) is non-perturbative in back-reaction and there seems to be a notion of finite-density universality in the IR: a certain scaling symmetry is emergent.

We intend to explore the above two observations with a somewhat different edge. The differences are manifold, of which a few highlighted ones are: $(i)$ We will merely emulate the back-reaction of flavours. Instead of constructing a completely stringy embedding, we will consider an {\it effective} gravity theory, where the matter source is described by a Dirac-Born-Infeld action. $(ii)$ We will forcefully turn the dilaton off, that is motivated essentially on the grounds of simplicity. This enforcement is certainly correlated to our inherently {\it bottom up} approach. $(iii)$ We will consider space-filling or partially space-filling Brane sources, and in the latter case with convenient smearing along the transverse directions. This, for our purpose, means that we will use a DBI-action of the same dimension as the gravity action. For this article, we will not consider any Wess-Zumino term. The idea of treating flavour back-reaction in a so-called {\it bottom up} model is not new, a large body of literature already exists exploring various aspects of QCD-like features, see {\it e.g.}~\cite{Arean:2013tja, Alho:2013hsa, Alho:2015zua, Jarvinen:2015ofa, Rougemont:2015oea, Drwenski:2015sha, Jarvinen:2015qaa, Gursoy:2016ofp}.

Clearly, the subsequent results that we obtain and further analyze are not immune to a possible lack of an UV-complete description, {\it i.e.}~we will not be able to clearly rule in or rule out a stringy embedding of everything that we observe. In this article, we are motivated by the somewhat universal and, perhaps with some literary freedom, the {\it attractor-type} behaviour of an IR scaling-symmetric (technically speaking, the Hyperscaling-violating Lifshitz) geometry obtained in \cite{Faedo:2014ana}. Thus persuaded, we will consider turning on two types of bulk gravitational fields that presumably correspond to, {\it via} the Gauge-String duality, a non-zero density (or, chemical potential) and a constant magnetic field. Both these correspond to relevant deformations of an UV CFT of certain dimensions, and the deformations are applied explicitly by the fundamental sector. As we consider the gravitational back-reaction by solving the resulting Einstein equations (along with a Maxwell-type one), we observe that, the deep infrared receives a qualitative correction. This correction, in an appropriate sense, is inherently {\it non-perturbative} in that a simple $N_f/ N_c$ correction is unlikely to yield the same.

The solutions that we obtain are of the following type: Starting with an AdS$_{d+1}$-dimensional UV, the density driven IR is given by an AdS$_{2} \times {\mathbb R}^{d-1}$. On the other hand, the magnetically driven IR turns out to be an AdS$_{d-1} \times {\mathbb R}^{2}$. We are, however, unable to find an analytical solution when both density and magnetic deformations are present; should an analytical solution exist, it is certainly not of scaling type. Note that, in both cases one turns on a bulk two-form field. In the density-driven case, the directions parallel to the Hodge dual of the two-form decouples from the dynamics; on the other hand, in the magnetically driven case, the directions parallel to the two-form do.

The emergence of an effective AdS$_2$, or an AdS$_{d-1}$ IR is not new. Similar physics is observed in {\it e.g.}~taking the near-horizon limit of an extremal Reissner-Nordstrom black hole, and in the solutions described in {\it e.g.}~\cite{D'Hoker:2009mm}. An AdS$_2$ has also been obtained in \cite{Alho:2013hsa}, from a {\it bottom up} construction of Veneziano limit, in {\it e.g.}~\cite{Hartnoll:2009ns, Pal:2012zn, Tarrio:2013tta} within the context of Gauge-Gravity duality, and earlier in {\it e.g.}~\cite{Cataldo:1999wr, Fernando:2003tz, Dey:2004yt} from a purely gravitational perspective, with an action similar to the one that we consider.\footnote{We thank Javier Tarr\'{i}o for pointing out these references to us.} Our work is along the lines of these earlier works, in which we explore this AdS$_2$ from a different perspective and with a complementary analysis, to {\it e.g.}~emphasize the non-perturbative nature of the IR. Moreover, we also obtain anisotropic solutions, which have not previously appeared in this context. On the other hand, compared to \cite{D'Hoker:2009mm}, there is another important physical difference: The IR is fundamental matter dominated, be it density or the magnetic field. We can equivalently state that our oversimplified model is sufficient to capture these features, which are nonetheless present in more rigorous {\it top down} stringy constructions. Thus, one perhaps does not need to resort to a precise stringy construction for addressing sufficiently general issues.\footnote{We should mention that the IR CFTs that emerge in our model, may as well remain in a more involved construction\cite{Faedo:2014ana}, if the dilaton vanishes at this point. In general though, including a non-trivial dilaton is more generic. We are currently exploring this and other possibilities.}

We also construct explicit flows to the corresponding IR CFTs, which are only numerical. It should be possible to construct a perturbative solution around each CFTs, or the AdS-fixed points. Already, the leading order perturbation, which we explicitly perform for each case, the corrections encode crucial information about the deformation, {\it e.g.}~the dimension of the corresponding operator. Treating the example of $5$-bulk dimensions, we observe that density perturbation is more relevant towards the IR. This is further corroborated by the linearized analysis near the IR fixed points: {\it via} a density perturbation around the magnetically driven AdS$_3 \times {\mathbb R}^2$ and a magnetic perturbation around the density-driven AdS$_2\times {\mathbb R}^3$ solutions. While the former is a relevant deformation, the latter is logarithmic. Towards the IR, this logarithmic divergence can simply be tamed by introducing an event horizon. Therefore, in the limit of a small magnetic field, the deep IR is dominated by a (thermal) AdS$_2$ and a corresponding asymptotic solution can be constructed. The AdS$_2$, on the other hand, will be drastically modified at the UV --- a property usual to AdS$_2$--gravity. See {\it e.g.}~\cite{Almheiri:2014cka} for a general analysis of back-reaction in AdS$_2$ from a different perspective. It would be interesting for us to understand and explore the flow to the AdS$_2$ further, in view of the current interests in AdS$_2$/CFT$_1$\cite{Jensen:2016pah, Maldacena:2016upp, Engelsoy:2016xyb}.

In carrying out the linearized analysis, we observe the following: The scale of back-reaction and the scale of conformal symmetry breaking are distinct. While the back-reaction always appears as a power law correction, the breaking of conformal symmetry is only perceived as the appearance of a log-term, {\it i.e.}~by inducing a conformal anomaly\cite{Skenderis:2002wp}. For example, conformal symmetry breaking seems to happen at a different scale that is closer to the {\it e.g.}~AdS$_3 \times {\mathbb R}^2$ fixed point, than the back-reaction scale.

We also find anisotropic solutions. For example, when a density is turned on, encoded in the gauge field $F = A_t'(r) dt \wedge dr$, we find an AdS$_2 \times$EAdS$_{d-1}$ geometry. Similarly, an AdS$_{d-1}\times$EAdS$_2$ solution exists with the two-form $F = dA$, where $A = A_y(x) dy$. Interestingly, such solutions also exist with the unflavoured action: Einstein gravity with a negative cosmological constant.\footnote{We are not aware whether this observation has been manifestly presented before. This generalizes to AdS and EAdS of various dimensions.} The main difference between the flavoured and the unflavoured cases is: in the former the curvature scales for the AdS and the EAdS can be arbitrary, whereas for the latter these two scales are locked. It is also straightforward to check that one can trivially introduce event horizon in these geometries. In the limit of  vanishing event horizon, {\it i.e.}~vanishing temperature in the dual field theory, AdS$_2 \times$EAdS$_{d-1}$ and AdS$_{d-1}\times$EAdS$_2$ are related simply by an analytic continuation. This is expected, since the unbroken Lorentz invariance allows us to trade freely between the (bulk) electric and magnetic configurations.

The case of $d=3$ is special, due to a (bulk) electric-magnetic duality (S-duality). In this case, contrary to the general story, an analytic scaling solution exists with both density and magnetic fields turned on. The anisotropic solution, in this case, is characterized by the usual scaling in the radial coordinate, along with a scaling in $y$ with a shift in $x$ direction. This is irrespective of the ``electric" or the ``magnetic" nature of the gauge field.

In one dimension higher, $d=4$, the features are more generic. The density driven phase singles out an AdS$_2$ and the decoupled $3$-manifold can be either ${\mathbb R}^3$ or an EAdS$_3$. The latter is an anisotropic solution of Bianchi type-V. In the latter, a shift in the $x$-direction along with rescaling in $y$ and $z$-directions constitute the corresponding symmetry. These are the only homogeneous and anisotropic solutions within the scaling ansatz.

In order to make connection with the existing literature, we note that in \cite{Iizuka:2012iv, Iizuka:2012pn, Kachru:2013voa} various anisotropic solutions of different Bianchi types were found within Einstein gravity with negative cosmological constant and a massive Proca field. This was also a {\it bottom up} or a {\it phenomenological} approach, in which the mass of the gauge field was treated a free parameter in the theory. In the limit of vanishing mass, no anisotropic solution survives. The putative dual field theory, in these cases, does not have any fundamental matter; the only degrees of freedom are adjoint fields. Thus, one can only switch on a density or a magnetic field in the adjoint sector. It is interesting to note that, with fundamental matter, the qualitative physics remains somewhat similar, {\it e.g.}~the effective dimensional reduction with a magnetic field, resulting from a frozen dynamics at the lowest Landau level.\footnote{With fundamental matter in the probe limit, this effective dimensional reduction is often thought to be responsible for the breaking of chiral symmetry as observed in various holographic models in {\it e.g.}~\cite{Filev:2007gb, Albash:2007bk, Erdmenger:2007bn, Bergman:2008sg, Johnson:2008vna, Alam:2012fw}.} It would be revealing to demonstrate this phenomenon in a suitable weakly coupled field theory, which we leave for a future work.

We briefly discuss the case of partially-filling brane sources. In this case, the DBI source has a reduced dimensionality compared to the one in which Einstein gravity is defined. This corresponds to introducing the fundamental matter sector as defects in a system of adjoints. To simplify the problem, we also smear the partially-filling branes along the transverse directions, thereby reducing the problem to unknown functions of only one, namely the radial, variable. It turns out, however, that within the scaling ansatz we find AdS$_{p+1} \times {\mathbb R}^{d-p}$ solutions, which are non-perturbative in back-reaction. The back-reacting brane is $(p+1)$-dimensional, living in a $(d+1)$-dimensional geometry. These geometries are purely coloured and flavoured, with no additional fields turned on. We do not find any analytical solution with a density or a magnetic field, in these cases.

This article is divided in the following sections: In section two, we introduce the action and explicitly write down the corresponding equations of motion. We discuss various solutions, in details, for the special case of $d=3$ in section 3. Subsequently, we comment on the general case in the next section. A detailed analysis, including a discussion of the dimension of various operators corresponding to the density and the magnetic field, from the perspective of various fixed points, is discussed in section 5, with the sufficiently general example in $d=4$. Finally, we offer a few comments on the partially-filling brane sources in the next section.

\section{The Action and the EOMs}

Our starting point is the following action:
\begin{eqnarray}
&& S_{\rm full} = S_{\rm gravity} + S_{\rm DBI} \ , \label{act1} \\
&& S_{\rm gravity} = \frac{1}{2 \kappa^2} \int d^{d+1} x \sqrt{- {\rm det} g} \ \left(R - 2 \Lambda \right) \ , \label{act2} \\
&& S_{\rm DBI} = - \tau \int d^{d+1} x \sqrt{- {\rm det} \left( g + F \right) } \ . \label{act3}
\end{eqnarray}
Here $S_{\rm gravity}$ represents Einstein-gravity that is typically dual to the adjoint sector of a gauge theory and $S_{\rm DBI}$ corresponds to the action of a brane that is dual to the fundamental sector of the field theory. Also, $\kappa$ represents the Newton's constant, $\tau$ represents the ``brane tension". The field $F$ is a $U(1)$-gauge field living on the brane. In the limit of small fields, $S_{\rm DBI}$ reduces to a simple Maxwell term, $S_{\rm Maxwell}$. An AdS-solution is obtained if $\Lambda = - d (d-1)/ 2 L^2$, where $L$ represents the radius of AdS.

The equations of motion resulting from the variation of the action are:
\begin{eqnarray}
&& R_{\mu\nu} - \frac{1}{2} \left( R - 2 \Lambda \right) g_{\mu\nu} = T_{\mu\nu} \ ,  \label{ein} \\
&& \partial_\mu \left( \sqrt {-{\rm det} \left( g + F \right)} \ \A^{\mu \nu} \right) = 0 \ , \label{max}
\end{eqnarray}
where 
\begin{eqnarray}
&& \A^{\mu \nu} = - \left( \frac{1}{g+F} \cdot F \cdot \frac{1}{g- F} \right)^{\mu\nu} \ , \label{anti} \\
&& T^{\mu\nu} = \frac{\kappa^2 \tau}{\sqrt{- {\rm det} g}} \left( \frac{\delta S_{\rm DBI}}{\delta g_{\mu\nu}} + \frac{\delta S_{\rm DBI}}{\delta g_{\nu\mu}} \right) = - \left( \kappa^2\tau \right) \frac{\sqrt{- {\rm det} \left( g+F \right)}}{\sqrt{ - {\rm det} g}} \S^{\mu\nu} \ , \\
&& \S^{\mu\nu} = \left( \frac{1}{g+F} \cdot g \cdot \frac{1}{g- F} \right)^{\mu\nu} \label{sym} \ .
\end{eqnarray}
In calculating the above, $\A^{\mu\nu}$ or $\S^{\mu\nu}$ can be evaluated by simply treating $g$ and $F$ as matrices, and then using the formulae in (\ref{anti}), (\ref{sym}).

To proceed further, we begin by fixing a dimension. For reasons of convenience, $d=3$ is a good choice: below this, gravity is non-dynamical and everything is essentially encoded within diffeomorphisms. Moreover, for $d=3$, the putative dual field theory is $(2+1)$-dimensional, and thus it can support a finite density, as well as a non-vanishing magnetic field along the field theory directions. We will, in due course of our discourse, discuss the physics in various dimensions.

\section{The Ansatz and the Solutions: $d=3$}

We begin our discussion in $d=3$, {\it i.e.}~in four bulk dimensions. We will discuss the generalizations afterwards. To warm up to the cause, let us start with the following ansatz:
\be 
ds^2 = -g_{tt}(r) dt^2 + g_{rr}(r) dr^2 + g_{xx}(r) dx^2 + g_{yy}(r) dy^2 \ ,
\ee
where the metric data $\{ g_{tt}(r), g_{rr}(r), g_{xx}(r), g_{yy}(r)\}$ are functions of the radial coordinate $r$ only. Now, we will discuss two distinct cases, in which we excite a gauge field in the DBI-sector that corresponds to a bulk electric field and a bulk magnetic field, respectively. These will be designed to, subsequently, correspond to a non-vanishing density (or a non-vanishing chemical potential) and a non-vanishing magnetic field in the conjectural dual field theory. We duly refer to these two cases as ``electric" and ``magnetic".

\subsection{The Electric Case}

We will discuss two inequivalent solutions in this section. The distinction lies in the behaviour of the metric.

\subsubsection{The ${\rm AdS}_2 \times \R^2$ Solution}

Let us begins with the following gauge-field ansatz:
\be \label{elecgauge}
A_{\mu} = \{A_t(r), 0,0,0 \} \ , 
\ee
and work with the following scaling-ansatz for the metric coefficients and the gauge field:
\be \label{elcmet}
\begin{split}
g_{tt}(r) = r^{\alpha},~g_{rr}(r) = r^{\beta},~g_{xx}(r)= g_{yy}(r) = r^{\delta} ~~\text{and}~~ A_t(r) = Q_e r^{\alpha_1} \ .
\end{split}
\ee
In what follows, we will explicitly discuss the strategy to obtain exact scaling-type solutions, that we use repeatedly in this article. In later sections, however, we will be terse.

First, the equations of motion for the gauge field becomes:
\be \label{eomgauge}
\frac{\alpha_1  Q_e r^{\alpha_1 +\delta -1} \left(r^{\alpha +\beta +2} (\alpha -2 \alpha_1 +\beta -2 \delta +2)+2  \delta\alpha_1 ^2  Q_e^2 r^{2 \alpha_1 }\right)}{2 \left(r^{\alpha +\beta +2}-\alpha_1 ^2 Q_e^2 r^{2 \alpha_1 }\right)^{3/2}} = 0 \ .
\ee
The $tt, rr, xx$ components of the Einstein's equation become:
\be \label{eomgrav}
\begin{split} 
-4 \Lambda -\frac{4 \kappa ^2 \tau  r^{\frac{1}{2} (\alpha +\beta +2)}}{\sqrt{r^{\alpha +\beta +2}-\alpha_1^2 Q_e^2 r^{2 \alpha_1}}}+\delta  (2 \beta -3 \delta +4) r^{-\beta -2}&=0 \ , \\
2 \alpha  \delta +\delta ^2+4 r^{\beta +2} \left(\Lambda +\frac{\kappa ^2 \tau  r^{\frac{1}{2} (\alpha +\beta +2)}}{\sqrt{r^{\alpha +\beta +2}-\alpha_1^2 Q_e^2 r^{2 \alpha_1}}}\right) &= 0 \ , \\
r^{\alpha } \left(\alpha ^2+\alpha  ( \delta-\beta -2)-(\beta +2) \delta +\delta ^2+4 \Lambda  r^{\beta +2}\right)& \\
+4 \kappa ^2 r^2 \tau  \sqrt{r^{\alpha +\beta -2} \left(r^{\alpha +\beta +2}-\alpha_1^2 Q_e^2 r^{2 \alpha_1}\right)} &= 0 \ ,
\end{split}
\ee
respectively.

It can now be seen from eq.\eqref{eomgauge}, that, for having a non-trivial scaling solution we must choose
\be
\alpha_1 = {\alpha +\beta + 2 \over 2} \ , 
\ee 
 which, in turn and to solve eq.\eqref{eomgauge}, requires
\be 
\delta = 0 \ .
\ee
With the above choices, the Einstein equations are solved by
\be \label{arbitalphasol1}
\beta = -2 \ ,~~ \Lambda = -{1 \over Q_e^2} \ ,~~ \tau = {\sqrt{4- Q_e^2 \alpha^2 } \over 2 Q_e^2 \kappa^2 } \ ,
\ee 
At this point we note the following: In the above equation, $\Lambda$ and $\tau$ define a bulk theory and can take any value. Given these, $Q_e$ and $\alpha$, which are integration constants of the particular solution, can be solved for using the above relations. In all subsequent cases, we write similar equations. These are to be interpreted as determining the integration constants, {\it i.e.}~$Q_e$ and $\alpha$, in terms of the parameters of the theory, {\it i.e.}~$\Lambda$ and $\tau$.  

Observe that, in the final solution, $\alpha$ remains undetermined:
\be \label{arbitalphamet1}
ds^2 = -r^{\alpha} dt^2 +{dr^2 \over r^2} + dx^2 + dy^2 \ .
\ee
The reason is that we are working in units where an overall length scale is set to unity. In other words, one can start from the metric in eq.\eqref{arbitalphamet1} and perform a coordinate transformation:
\be 
r= \tilde{r}^{\alpha \over 2}~~\text{and} ~~ t= {2 \over \alpha} \tilde{t} \ ,
\ee
such that we obtain 
\be 
ds^2 ={4 \over \alpha^2} \left[-\tilde{r}^2 d\tilde{t}^2 +{d\tilde{r}^2 \over \tilde{r}^2}\right] + dx^2 + dy^2.
\ee
The above clearly factors out an overall numerical constant. Basically, we obtained an AdS$_2 \times \R^2$ solution, in which the AdS$_2$ length-scale is determined by $\alpha$. This length can always be factored out by rescaling the coordinates $x,~y$ and, hence, has no physical consequence. Certainly, we can work in units where $\alpha=2$, {\it i.e.}~choosing the AdS$_2$ radius to be unity, and we obtain the solution:
\be 
\alpha =2 \ ,~~ \alpha_1 = 1 \ ,~~ \beta = -2 \ ,~~ \Lambda = -{1 \over Q_e^2} \ ,~~ \tau = {\sqrt{1- Q_e^2} \over  Q_e^2 \kappa^2 } \ . \label{adsqe}
\ee
We end this section with a comment. Note that, by looking at (\ref{adsqe}), it na\'{i}vely seems that a well defined $\tau = 0$ limit exists and it is obtained by setting $Q_e = 1$. This, however, is untrue. Going back to the original equation in (\ref{eomgauge}), it is straightforward to check that setting $Q_e = 1$ also exacts the denominator to vanish, thereby annulling the subsequent analysis, altogether. Alternatively, it can also be checked explicitly that AdS$_2 \times \R^2$ does not extremize the action in (\ref{act1}), when $\tau = 0$. We can arrange $Q_e$ approach as close to unity as possible, subsequently tuning $\tau \to 0$. This, however, is non-perturbative, since $Q_e$ needs to be tuned to the maximum allowed value. Thus, the solution is non-perturbative in back-reaction.\footnote{We will discuss this in some details, in section \ref{pertnpert}. One can look for the case when AdS$_2$ and $\R^2$ come with separate length-scales, denoted by $L_1$ and $L_2$, such that $ds_{{\rm AdS}_2}^2 = L_1^2 \left( - r^2 dt^2 + dr^2 / r^2 \right) $, and $ds_{\R^2}^2 = L_2^2 \left( dx^2 + dy^2 \right)$. As expected, the solution is $L_2$-independent, since it merely rescales the spatial coordinates. The AdS-radial scale, however, sets the maximum value of $Q_e$, {\it via} the following relation:
\begin{eqnarray}
\tau = \frac{\sqrt{L_1^2 - Q_e^2}}{Q_e^2 \kappa^2} \ .
\end{eqnarray}
}

\subsubsection{The ${\rm AdS}_2 \times {\rm EAdS}_2$ Solution}

With the same gauge field, there is another exact solution which we discuss below. Now, the metric and gauge field scaling-ansatz goes as: 
\be \label{elcmet1}
\begin{split}
&g_{tt}(r) = L_1 r^{\alpha} \ ,~g_{rr}(r) = L_1  r^{\beta} \ ,~g_{xx}(r)= L_2  r^{\delta} \ ,~ g_{yy}(r,x) =L_2 e^{-2 x}  r^{\delta} \ , \\ 
& ~~\text{and}~~ A_t(r) = Q_e r^{\alpha_1} \ .
\end{split}
\ee
The difference from the previous case clearly lies in the explicit $x$-dependence of the $g_{yy}$-component, and hence, the geometry is homogeneous, but not isotropic. Note, also, that we have introduced two different length scales, $L_1$ and $L_2$. However, as we will see, only there ratio is physical.

It can be checked that there is solution of the following form:
\be
\alpha=-\beta = 2 \ ,~~ \delta  = 0 \ ,~~ \alpha_1 =1 \ , ~~ \Lambda = \frac{L_1 ^2-L_1L_2 -Q_e ^2}{L_2Q_e ^2} \ ,~~\tau = \frac{(L_2 -L_1 ) \sqrt{L_1 ^2-Q_e ^2}}{\kappa ^2 L_2Q_e ^2} \ . \label{eadsqe}
\ee
It is clear that $L_1,~L_2$ appears only in the dimensionless combination of $\left( L_1 / L_2 \right)$, {\it i.e.}~the ratio of the two radii of ${\rm EAdS}_2$ and ${\rm AdS}_2$ geometries.\footnote{That the geometry in (\ref{elcmet1}) corresponds to an AdS$_2\times$EAdS$_2$ is best seen using the following coordinate change: $x = \log u$.}

Interestingly, note that in the case $L_1 = L_2$, the DBI part of the action decouples from the system, since $\tau = 0$. This suggests that there is a similar ${\rm AdS}_2 \times {\rm EAdS}_2$ geometry with Einstein gravity and a negative cosmological constant:
\be
\alpha=-\beta = 2 \ ,~~ \delta  = 0 \ ,  ~~ \Lambda = -\frac{1}{L_1} \ ,~~\tau =0 \ ,
\ee
that can also be explicitly checked. On the contrary, this is not true for the ${\rm AdS}_2 \times \R^2$ solution, for which a non-vanishing contribution from DBI is necessary.

It is worth noting that the two solutions discussed above, {\it i.e.}~$ {\rm AdS}_2 \times \R^2$ and $ {\rm AdS}_2 \times {\rm EAdS}_2 $, are the only two possible homogeneous, but not necessarily isotropic, solutions within the scaling-ansatz.

\subsection{The Magnetic Case}

As before, we will also discuss two inequivalent solutions in this section. We will also present some of the details in this section.

\subsubsection{The ${\rm AdS}_2 \times \R^2$ Solution}

Now, consider the following gauge field:
\be 
A_{\mu} = \{0, 0,A_x(y),0 \} \ ,
\ee
with the following scaling-ansatz for the metric coefficients and the gauge field:
\be 
\begin{split}
g_{tt}(r) = r^{\alpha} \ ,~g_{rr}(r) = r^{\beta} \ ,~g_{xx}(r)= g_{yy}(r) = r^{\delta} ~~\text{and}~~ A_x(y) = Q_m y \ .
\end{split}
\ee
The equation for the gauge field is identically satisfied. The Einstein's equations yield:
\be 
\begin{split}
& \kappa ^2 \tau  \sqrt{Q_m^2+r^{2 \delta }} r^{\beta -\delta }+\Lambda  r^{\beta }+\frac{\delta  (2 \alpha +\delta )}{4 r^2} = 0 \ , \\
& -4 \Lambda -4 \kappa ^2 \tau  r^{-\delta } \sqrt{Q_m^2+r^{2 \delta }}+\delta  (2 \beta -3 \delta +4) r^{-\beta -2} = 0 \ , \\
& \frac{\kappa ^2 \tau  r^{2 \delta }}{\sqrt{Q_m^2+r^{2 \delta }}}+\frac{1}{4} \left(\alpha ^2+\alpha  (-\beta +\delta -2)+\delta  (-\beta +\delta -2)\right) r^{-\beta +\delta -2}+\Lambda  r^{\delta } = 0 \ .
\end{split}
\ee
One solution of the equations above is:
\be 
\delta=0 \ ,~~\beta=-2 \ ,~~\Lambda = -\frac{\alpha ^2 \left(Q_m^2+1\right)}{4 Q_m^2} \ ,~~\tau = \frac{\alpha ^2 \sqrt{Q_m^2+1}}{4 \kappa ^2 Q_m^2} \ .
\ee
Once again, $\alpha$ remains undetermined, and we get:
\be 
ds^2 = -r^{\alpha} dt^2 +{dr^2 \over r^2} + dx^2 + dy^2 \ .
\ee
Thus, we get a similar ${\rm AdS}_2 \times \R^2$ solution with the choice $\alpha=2$,
\be 
\Lambda = -\frac{\left(Q_m^2+1\right)}{ Q_m^2} \ ,~~\tau = \frac{\sqrt{Q_m^2+1}}{ \kappa ^2 Q_m^2} \ . \label{adsqm}
\ee
There is, however, an important difference between the solution described in (\ref{adsqe}) and the one in (\ref{adsqm}). While the one in (\ref{adsqe}) has a well-defined $\tau \to 0$ limit, the above solution does not. The easiest way to see this is to express $Q_m$ in terms of $\tau$, in the limit $\tau \to 0$:
\be
Q_m^ 2  = \frac{\alpha^2}{16 \kappa^4 \tau^2 } + 1 + \O(\tau^3) \ ,
\ee
which is singular.

\subsubsection{The ${\rm AdS}_2 \times {\rm EAdS}_2$ Solution}

As before, we also get the ${\rm AdS}_2 \times {\rm EAdS}_2$ solution. The corresponding metric functions and the gauge field are:
\be \label{magmet1}
\begin{split}
&g_{tt}(r) = L_1 r^{\alpha} \ ,~g_{rr}(r) = L_1  r^{\beta} \ ,~g_{xx}(r)= L_2  r^{\delta} \ ,~ g_{yy}(r,x) =L_2 e^{-2 x}  r^{\delta} \ ,\\
& A_y(x) = Q_m e^{-\alpha_1 x} \ .
\end{split}
\ee
The corresponding solution is obtained by
\be \label{eadsqm}
\begin{split}
& \alpha=-\beta = 2 \ ,~~ \delta  = 0 \ ,~~ \alpha_1 =1 \ , ~~ \Lambda = -\frac{-L_1 L_2+L_2^2+Q_m^2}{L_1 Q_m^2} \ , \\
& \tau = \frac{(L_2-L_1) \sqrt{L_2^2+Q_m^2}}{\kappa ^2 L_1 Q_m^2} \ . 
\end{split}
\ee
As before, in the limit $L_1=L_2$, the DBI sector decouples and this can be obtained as a solution of Einstein gravity with a negative cosmological constant. Once again, $\tau \to 0$ limit is singular, unless we also tune $L_1 \to L_2$, and the above solution cannot be obtained treating the DBI backreaction perturbatively.

Before discussing the general case, let us make an explicit connection between the electric and the magnetic solutions that are related by an S-duality. It is straightforward to check that, under the following map:
\begin{eqnarray}
\varphi_{\rm S-dual} : Q_m^2 \rightarrow \frac{L_2^2 Q_e^2}{L_1^2 - Q_e^2} \ ,
\end{eqnarray}
the corresponding solutions are mapped as:
\begin{eqnarray}
\varphi_{\rm S-dual} : (\ref{adsqe}) \rightarrow (\ref{adsqm})  \ , \quad \varphi_{\rm S-dual} : (\ref{eadsqe}) \rightarrow (\ref{eadsqm}) \ .
\end{eqnarray}
%

\subsection{The Electric-Magnetic Case}

As a natural continuation of the above results, let us now explore the gauge field with both magnetic and electric components. The gauge field and the metric data are:
\be
\begin{split}
& A_{\mu} = \{A_t(r), 0,A_x(y),0 \} \ ,   \text{with}~~ A_x(y) = Q_m y \ , ~~A_t(r) = Q_e r^{\alpha_1} \ , \\
& g_{tt}(r) = L_1 r^{\alpha} \ ,~g_{rr}(r) = L_1 r^{\beta} \ ,~g_{xx}(r)= g_{yy}(r) = L_2 r^{\delta}  \ . 
\end{split}
\ee
The AdS$_2 \times \R^2$ solution is simply obtained to be:
\be 
\begin{split}
& \alpha_1 = {\alpha \over 2} = 1 \ ,~~ \delta =0 \ ,~~\beta = -2 \ ,~~ \Lambda = -\frac{L_1 \left(Q_m^2+ L_2^2\right)}{L_2 ^2 Q_e^2 + L_1^2 Q_m^2} \ , \\
& \tau = \frac{ L_2 \sqrt{\left(Q_m^2 + L_2^2\right) \left(L_1^2 - Q_e^2 \right)}}{ \kappa ^2 \left(L_2^2 Q_e^2 + L_1^2 Q_m^2\right)} \ .
\end{split}
\ee
Clearly, $\tau \to 0$ limit is smooth if we tune $Q_e \to 1$, but it is singular if we hold $Q_e \not = 1$ fixed.

On the other hand, the AdS$_2 \times$EAdS$_2$ solution can be characterized by the following data: First, we write down the ansatz for the metric and the gauge field as:
\be
\begin{split}
& A_{\mu} = \{A_t(r), 0,A_x(y),0 \} \ , \\
& g_{tt}(r) =  L_1 r^{\alpha} \ ,~g_{rr}(r) = L_1 r^{\beta} \ ,~g_{xx}(r)=L_2 r^{\delta} \ ,~g_{yy}(r,x) =  L_2 e^{-2x} r^{\delta}  \\
& \text{and}~~ A_y(x) = Q_m  e^{-\alpha_1 x} \ , ~~A_t(r) = Q_e r^{\alpha_2} \ .
\end{split}
\ee
The solution is given by
\be 
\begin{split}
& \alpha_1 = 1 \ ,~~ \alpha_2 = 1 \ ,~~\delta =0 \ ,~~\beta = -2 \ ,~~\alpha=2 \ , \\
& \Lambda = \frac{L_1^2 L_2-L_1 \left(L_2^2+Q_m^2\right)-L_2 Q_e^2}{L_1^2 Q_m^2+L_2^2 Q_e^2} \ , \\
& \tau = \frac{(L_2-L_1) \sqrt{(L_1-Q_e) (L_1+Q_e) \left(L_2^2+Q_m^2\right)}}{\kappa ^2 \left(L_1^2 Q_m^2+L_2^2 Q_e^2\right)} \ .
\end{split}
\ee
As before, $L_1 = L_2$ limit exists, corresponding to $\tau = 0$, in which the DBI sector decouples.

\subsection{Perturbative or Non-perturbative} \label{pertnpert}

In this section, we will formally define and subsequently classify the already discussed solutions as perturbative or non-perturbative in back-reaction. The action in (\ref{act1})-(\ref{act3}) has two paramaters: $\Lambda$ and $\kappa^2\tau$. The solutions are characterized by four other parameters: $Q_e, Q_m, L_1, L_2$, which are related to the parameters of the action. Each corresponding solution also comes with a regime of validity for these parameters. Now, we will define a solution as {\it perturbative}, provided: $(i)$ One can tune $\kappa^2 \tau \to 0$ within the regime of validity for various parameters characterizing the solution, $(ii)$ the same solution is obtained by setting $\kappa^2 \tau = 0$, which corresponds to the zeroth order result. A solution that violates either of these two conditions, will be characterized as non-perturbative.

Now, from (\ref{adsqe}) we get:
\begin{eqnarray}
Q_e^2 = \frac{2}{1 + \sqrt{4 \kappa^4\tau^2 + 1}} \quad \implies \quad Q_e < 1 \ .
\end{eqnarray}
Since we cannot reach $Q_e = 0$, it already violates condition $(ii)$ above. On the other hand, for the solution in (\ref{adsqm}) the limit $\kappa^2 \tau \to 0$ is singular and thus violates condition $(i)$ above. Thus, both solutions are non-perturbative. Furthermore, even though the $\kappa^2\tau \to 0$ limit seems to have distinct behaviours in (\ref{adsqe}) and (\ref{adsqm}), we argue below that this is not the case according to our criteria set above. Towards that, note the following:
\begin{eqnarray}
&& Q_e^2 = 1 - \kappa^4 \tau^2 + \O\left( \kappa^8\tau^4 \right)  \ , \\
&& Q_m^2 = \frac{1}{\kappa^4 \tau^2} + 1 - \kappa^4 \tau^2 + \O\left( \kappa^8\tau^4 \right)  \ ,
\end{eqnarray}
along with the corresponding regimes of validity: $ 0 < Q_e < 1$ and $0 < Q_m < \infty$. In both cases, $\kappa^2\tau \to 0$ limit is connected to the $Q_e \to 1$ or $Q_m \to \infty$ limit, respectively; while setting $\kappa^2\tau = 0$ demands us to set $Q_e =1 $ or $Q_m =\infty$, respectively. These features are identical in both solutions.

Now, let us consider (\ref{eadsqe}). Any solution characterized by $L_1 / L_2 \not = 1$ cannot be obtained with $\kappa^2 \tau = 0$, even though at precisely $L_1 = L_2$, the solution exists with $\kappa^2 \tau =0$. Thus, since condition $(ii)$ is violated, the solution in (\ref{eadsqe}) is also non-perturbative. A similar conclusion can be drawn for the solution in (\ref{eadsqm}). When both $Q_e$ and $Q_m$ are present, one can expand in $\kappa^2\tau$ keeping either of these fixed, and arrive at a similar conclusion. Thus, in brief, all solutions are non-perturbative in back-reaction.

\section{The Ansatz and the Solutions: General Dimensions}

In view of the AdS$_2 \times \R^2$ solution that we obtained with both electric and magnetic sources in the previous section, we will now comment on the higher dimensional generalization. The generalization turns out to be rather simple and intuitive: With a purely electric field, in $(d+1)$-bulk dimensions, {\it i.e.}~when the boundary field theory is $d$-dimensional, there is an AdS$_2 \times \R^{d-1}$ solution. With a magnetic field, however, the analogous exact solution is AdS$_{d-1} \times \R^2$. For $d=3$, they are both AdS$_2 \times \R^2$, which we have explicitly obtained before. In the dual $d$-dimensional field theory, this implies that at non-vanishing density the IR-phase is always dominated by a $(0+1)$-dimensional CFT. On the other hand, if we couple the system with a constant magnetic field, then the IR-phase is dominated by a $(d-2)$-dimensional CFT.

In this section we briefly discuss explicit solutions. In the purely electric case, let us begin with the following scaling ansatz:
\be 
\begin{split}
 &~ A_t(r) = Q_e r^{\alpha_1} \ ,~~ A_r = A_i =0 \ , ~\text{for all}~ i = 1,\cdots ,d-1 \ . \\
& ds^2 = -g_{tt}(r) dt^2 + g_{rr}(r) dr^2 + g_{11}(r)  \sum_{i=1}^{d-1} dx_i^2 \ , \\
&g_{tt}(r) = L_1 r^{\alpha} \ ,~g_{rr}(r) = L_1 r^{\beta} \ ,~g_{11}(r) = L_2 r^{\delta} \ .
\end{split}
\ee
The corresponding AdS$_2 \times \R^{d-1}$ solution is given by
\be 
\alpha =2 \ ,~~ \alpha_1 = 1 \ ,~~ \beta = -2 \ ,~~ \Lambda = -{L_1 \over Q_e^2} \ ,~~ \tau = {\sqrt{L_1^2 - Q_e^2} \over  Q_e^2 \kappa^2 } \ . \label{adsqed}
\ee
On the other hand, we can consider a magnetic field, to be concrete in $(d+1)$-bulk dimensions, of the following form:
\be 
\begin{split}
&~ A_{x_1}(x_2) = Q_m x_2 \ ,~~ A_r = A_t =A_i = 0 \ , ~\text{for all}~ i = 3,\cdots ,d-4 \ .  \\
& ds^2 = -g_{tt}(r) dt^2 + g_{rr}(r) dr^2 +g_{11}(r) (dx_1^2 + dx_2^2) +g_{33}(r)  \sum_{i=3}^{d-4} dx_i^2~ , \\
& g_{tt}(r) = L_1 r^{\alpha} \ ,~g_{rr}(r) = L_1 r^{\beta} \ ,~g_{11}(r) = L_2 r^{\delta} \ ,~g_{33}(r) = L_1 r^{\sigma} \ .
\end{split}
\ee
To be concrete, we consider the example of $d=4$. In this case, we find an ${\rm AdS}_3 \times \R^{2}$ solution as given below:
\be 
\alpha =\sigma = 2 \ ,~~ \delta =0 \ ,~~\beta = -2 \ ,~~ \Lambda = - \frac{1}{L_1}\left( \frac{2 L_2^2}{Q_m^2} + 3\right) \ ,~~\tau = \frac{2 L_2 \sqrt{Q_m^2 + L_2^2}}{ L_1 \kappa ^2 Q_m^2} \ . \label{adsqmd}
\ee
It is now straightforward to check that, according to the criteria set in section \ref{pertnpert}, the above solutions are also non-perturbative in back-reaction.

\section{The Ansatz and the Solutions: $d= 4$} \label{sect:sold4}

Now we specifically consider a $(4+1)$-dimensional bulk. The action that we extremize remains the same as in eq.\eqref{act1}. We will also explore homogeneous, but anisotropic solutions. The AdS$_2 \times \R^3$ (electric) and AdS$_3\times \R^2$ (magnetic) solutions evidently exist and are given by (\ref{adsqed}) and (\ref{adsqmd}). Note that, in this case, both AdS$_3$ and AdS$_2$ appear in the IR, depending on the UV-deformation.

These solutions are already discussed as a part of the general story in $(d+1)$-bulk dimensions. We will, now, comment on the physics.  First, let us comment on the operators that we turn on at the UV --- that is described by a $(3+1)$-dimensional CFT --- corresponding to the bulk magnetic and the electric deformations. We will do this by performing a perturbative analysis around the AdS$_5$-asymptotics.

In the case of a magnetic field, it is straightforward to check that $F = Q_m dx^1 \wedge dx^2$ satisfies (\ref{max}) trivially, irrespective of the geometry; thus we need not concern with a perturbative solution for the gauge field. Assuming that $Q_m$ is ``small", equivalently expanding around the AdS$_5$-asymptotics, one can easily calculate corrections to the metric at the leading order in $Q_m^2$. Renaming $g_{11} = g_{yy}$ and $g_{33} = g_{xx}$, this yields:
\begin{eqnarray}
&& \Lambda = - \frac{6 + L_1 \kappa^2 \tau}{L_1} \ , \\
&& g_{tt} = L_1 r^2 \left(1 + \delta g_{tt} \right) \ , \quad g_{xx} = L_1 r^2 \left( 1 + \delta g_{xx}\right) \ , \quad g_{yy} = L_1 r^2 \left( 1 + \delta g_{yy} \right) \ , \\
&& g_{rr} = L_1 r^{-2} \left( 1 + \delta g_{rr}\right) \ ,
\end{eqnarray}
where
\begin{eqnarray}
&& \delta g_{tt} = Q_m^2 \left[ \frac{\alpha_t^{(1)}}{r^4} + \frac{\alpha_t^{(2)}}{r^4} \log\left( r \right) \right] \ , \quad \delta g_{xx} = Q_m^2 \left[ \frac{\alpha_x^{(1)}}{r^4} + \frac{\alpha_x^{(2)}}{r^4} \log\left( r \right) \right] \ , \\
&& \delta g_{yy} = Q_m^2 \left[ \frac{\alpha_y^{(1)}}{r^4} + \frac{\alpha_y^{(2)}}{r^4} \log\left( r \right) \right] \ , \quad \delta g_{rr} = Q_m^2 \left[ \frac{\alpha_r^{(1)}}{r^4} + \frac{\alpha_r^{(2)}}{r^4} \log\left( r \right) \right] \ , \\
\end{eqnarray}
with the following constraints:
\begin{eqnarray}
&& \alpha_x^{(2)} = \alpha_t^{(2)} \ , \quad \alpha_y^{(2)} = \alpha_t^{(2)} + \frac{\kappa^2\tau}{2 L_1} \ , \quad \alpha_r^{(2)} = - 4 \alpha_t^{(2)} - \frac{\kappa^2\tau}{L_1} \ , \\
&& \alpha_r^{(1)} = - \alpha_t^{(1)} + \alpha_t^{(2)} - \alpha_x^{(1)} - 2 \alpha_y^{(1)} + \frac{\kappa^2\tau}{3 L_1} \ .
\end{eqnarray}
Thus, na\'{i}vely, the deformation is characterized by four free parameters. Note, also, that the magnetic perturbation behaves like a relevant deformation (since it grows towards the IR), and corresponds to a (mass scaling) dimension $2$ operator.\footnote{Recall that, the asymptotic fall-off behaviour $\Phi \sim ()_1 \ r^{-\Delta} + ()_2 \ r^{\Delta - d}$ applies to the metric components as well\cite{Witten:1998qj}, where $\Phi$ is a generic bulk field.. Here $\Delta$ is the mass scaling dimension of the operator. In this case, with $d=4$, we get $\Delta = 4$, which is the correct dimension of a boundary stress-energy tensor. Of course, the stress-tensor has twice the dimension of the magnetic field.} The leading order correction also involves a logarithmic contribution, that encodes breaking of conformal symmetry associated with the explicit scale set by the magnetic field. Without any loss of generality, we can set $\delta g_{rr} = 0$. This specifically yields:
\begin{eqnarray}
\varepsilon + p_x + 2 p_y = \frac{1}{12} \frac{\kappa^2 \tau}{L_1} \ ,
\end{eqnarray}
with the following identifications:
\begin{eqnarray}
\varepsilon = \alpha_t^{(1)} \ , \quad p_x = \alpha_x^{(1)} \ , \quad p_y = \alpha_y^{(1)} \ . \label{eosmag}
\end{eqnarray}
See {\it e.g.}~equation (\ref{deltag2}). In the equation of state, given in (\ref{eosmag}), $\varepsilon$, $p_x$ and $p_y$ are energy, pressure parallel and perpendicular to the magnetic field, respectively -- as viewed in the dual field theory. The equation of state has a non-vanishing right hand side, which signals breaking of conformal invariance. Recall that, typically, $\kappa^2 \sim N_c^{-2}$ and $\tau \sim N_f N_c$. Thus, conformal invariance is broken at ${\cal O} \left( N_f / N_c\right) $, which is, intuitively, expected. All in all, the number of free parameters is reduced to two: the energy and the anisotropy in pressure.

For the bulk electric field, which is dual to turning on a density perturbation on the boundary CFT, a similar calculation can be done, and the result can succinctly be presented as:
\begin{eqnarray}
&& \Lambda = - \frac{6 + L_1 \kappa^2 \tau}{L_1} \ , \\
&& g_{tt} = L_1 r^2 \left(1 + \delta g_{tt} \right) \ , \quad g_{xx} = L_1 r^2 \left( 1 + \delta g_{xx}\right) \ , \quad g_{yy} = L_1 r^2 \left( 1 + \delta g_{yy} \right) \ , \\
&& g_{rr} = L_1 r^{-2} \left( 1 + \delta g_{rr}\right) \ ,
\end{eqnarray}
where
\begin{eqnarray}
&& \delta g_{tt} = \frac{\varepsilon}{r^4} + Q_e^2  \frac{\beta_t^{(1)}}{r^6} \ , \quad \delta g_{xx} = \frac{p_x}{r^4} + Q_e^2  \frac{\beta_x^{(1)}}{r^6} =  \delta g_{yy} \ , \quad \delta g_{rr} = Q_e^2 \frac{\beta_r^{(1)}}{r^6} \ , \\
&& {\rm and} \quad \partial_r A_t(r) = \frac{Q_e}{\sqrt{L_1}} \frac{1}{r^3} \ ,
\end{eqnarray}
with the following constraints:
\begin{eqnarray}
&& \beta_x^{(1)} = \beta_t^{(1)} - \frac{\kappa^2\tau}{6 L_1^2} = \beta_y^{(1)}  \ , \quad \beta_r^{(1)} = - 6 \beta_t^{(1)} + \frac{5 \kappa^2\tau}{6 L_1^2} \ . 
\end{eqnarray}
In the above, we have written down the large $r$-asymptotic solution, in which energy and pressure terms are leading compared to the density perturbation. The metric deformation here corresponds to the addition of a (mass) dimension $6$ operator, and thus the gauge field deformation corresponds to turning on a (mass) dimension $3$ operator. As before, setting $\delta g_{rr} = 0$, we completely specify all asymptotic data:
\begin{eqnarray}
\beta_t^{(1)} = \frac{5}{36} \frac{\kappa^2 \tau}{L_1^2} \ , \quad \beta_x^{(1)} = - \frac{1}{36} \frac{\kappa^2 \tau}{L_1^2} \ ,
\end{eqnarray}
which are also ${\cal O} \left( N_f / N_c\right)$.

Now that we have a basic understanding of the operators turned on at the UV-boundary, we would like to perform a similar analysis in the IR. This is particularly facilitated by the fact that the IR is also a CFT, either an $(1+1)$-dimensional or a $(0+1)$-dimensional one, depending on the magnetic or the density deformation, respectively. Now we want to comment on the physics when both deformations are present, in which we do not find any analytical scaling-type solution. However, we can certainly estimate --- as viewed from the respective CFT --- what operator is turned on at the AdS$_3$ and the AdS$_2$ fixed points, corresponding to a density and the magnetic deformations, respectively.

We begin with the AdS$_3$ fixed point. Recall that this solution is given by (see (\ref{adsqmd}))
\begin{eqnarray}
&& ds^2 = L_1 \left( - r^2 dt^2 + r^2 dx^2 + \frac{dr^2}{r^2} \right) + L_2 d\vec{y_2}^2 \ , \\
&& \Lambda = - \frac{3 Q_m^2 + 2 L_2^2}{L_1 Q_m^2 } \ , \quad \tau = \frac{2 L_2}{L_1 Q_m^2 \kappa} \sqrt{L_2^2 + Q_m^2} \ .
\end{eqnarray}
Now we consider the following linearization
\begin{eqnarray}
&& g_{tt} = L_1 r^2 \left(1 + \delta g_{tt} \right) \ , \quad g_{xx} = L_1 r^2 \left( 1 + \delta g_{xx}\right) \ , \quad g_{yy} = L_2 \left( 1 + \delta g_{yy} \right) \ , \\
&& g_{rr} = L_1 r^2 \left( 1 + \delta g_{rr}\right) \ , \quad F = Q_m dy_1 \wedge dy_2 + \delta F \ .
\end{eqnarray}
Without any loss of generality, we can choose $\delta g_{rr} = 0$. This yields:
\begin{eqnarray}
&& \delta g_{tt} = \frac{\varepsilon}{r^2} - Q_e^2 \, \gamma_t^{(1)} \frac{1}{r^2} - Q_e^2 \, \gamma_t^{(2)} \frac{\log r}{r^2} \ , \label{ads3denpert1} \\
&& \delta g_{xx} = \frac{p_x}{r^2} + Q_e^2 \, \gamma_x^{(2)}  \frac{\log r}{r^2} \ , \label{ads3denpert2} \\
&& \delta g_{yy} = Q_e^2 \, \gamma_y^{(1)} \frac{1}{r^2} \ , \label{ads3denpert3} \\
&& \delta F = \frac{Q_e}{r} dt \wedge dr \ , \quad {\rm with} \quad \varepsilon + p_x = 0 \ .\label{ads3denpert4}
\end{eqnarray}
The various constants are:
\begin{eqnarray}
&& \gamma_t^{(1)} = \frac{2 L_2^2 \left(Q_m^2 + L_2^2\right)}{Q_m^2 L_1^2\left(3 Q_m^2 + 4 L_2^2\right)} \ , \quad \gamma_t^{(2)} = \frac{ \left( Q_m^2 + L_2^2\right)}{Q_m^2  L_1^2} \ , \\
&& \gamma_x^{(2)} = \frac{Q_m^2 + L_2^2}{Q_m^2 L_1^2} \ , \quad \gamma_y^{(1)} = \frac{Q_m^4 + 3 Q_m^2 L_2^2 + 2L_2^4}{2 Q_m^2 L_1^2 \left(3 Q_m^2 + 4 L_2^2\right)} \ .
\end{eqnarray}
Clearly, the equation of state, given in equation (\ref{ads3denpert4}), remains unaffected. Also, both $g_{tt}$ and $g_{xx}$ receive a logarithmic correction sourced by the density deformation. This metric deformation, as viewed from the CFT$_2$ perspective, is relevant, has mass dimension $2$ (therefore, the density perturbation turns on an operator with dimension $1$) and grows towards the IR. The logarithmic correction is absent in $g_{yy}$, but the deformation is still relevant. The presence of the logarithm function is associated with the breaking of conformal invariance due to non-vanishing density. One can, thus, identify two natural length scales: one where conformal symmetry is broken, and the other where density begins dominating the IR. The former can be identified by setting ${\cal O} \left( Q_e^2 \, \gamma_t^{(2)} \frac{\log r}{r^2} \right) \sim {\cal O}(1)$, while the latter is located at ${\cal O} \left( Q_e^2 \, \gamma_t^{(1)} \frac{1}{r^2} \right) \sim {\cal O}(1)$. Thus, the density dominated phase appears at a scale much lower than the scale of breaking conformal invariance.

We now move on to discussing the other IR: AdS$_2 \times {\mathbb R}^3$. The corresponding solution is given by (see equation (\ref{adsqed}))
\begin{eqnarray}
&& ds^2 = L_1 \left( -r^2 dt^2 + \frac{dr^2}{r^2} \right) + L_2 \left( dx^2 + d\vec{y_2}^2 \right)  \ , \\
&& \Lambda = - \frac{L_1}{Q_e^2} \ , \quad \tau = \frac{\sqrt{L_1^2 - Q_e^2}}{Q_e^2 \kappa^2 L_1} \ .
\end{eqnarray}
As before, we write down the following linearization:
\begin{eqnarray}
&& g_{tt} = L_1 r^2 \left(1 + \delta g_{tt} \right) \ , \quad g_{xx} = L_1 r^2 \left( 1 + \delta g_{xx}\right) \ , \quad g_{yy} = L_2 \left( 1 + \delta g_{yy} \right) \ , \\
&& g_{rr} = L_1 r^2 \left( 1 + \delta g_{rr}\right) \ , \quad F = - Q_e dt \wedge dr + \delta F \ .
\end{eqnarray}
This yields:
\begin{eqnarray}
&& \delta g_{tt} = Q_m^2 \, \frac{2}{3} \frac{L_1^2 - Q_e^2}{L_2^2 Q_e^2} \ , \quad \delta g_{rr} = Q_m^2 \, \frac{1}{3} \frac{L_1^2 - Q_e^2}{L_2^2 Q_e^2} \ , \\
&& \delta g_{xx} = Q_m^2 \,  \frac{4}{3} \frac{L_1^2 - Q_e^2}{L_2^2 Q_e^2} \, \log r = - \delta g_{yy} \ , \\
&& \delta F = Q_m \, dy_1 \wedge dy_2 \ .
\end{eqnarray}
In writing the above, we have explicitly left out the homogeneous solutions that are explicitly given in equations (\ref{ads2fluc1})-(\ref{ads2fluc3}). Quite clearly, $g_{tt}$ and $g_{rr}$ are merely renormalized, while $g_{xx}$ and $g_{yy}$ receive logarithmic corrections. Viewed from a purely AdS$_2$ perspective, this growth destroys the AdS$_2$ asymptotic, as well as the AdS$_2$ IR. One simple way to protect the IR is to excite the mode in (\ref{ads2fluc2}), which corresponds to introducing an event horizon. Note that, in this case, the scale of breaking conformal invariance and the scale of magnetic domination are one and the same, obtained by setting ${\cal O} \left( \delta g_{xx} \right) \sim {\cal O}(1)$. Thus, it is likely that, an RG flow connects the AdS$_3\times {\mathbb R}^2$ UV to the AdS$_2 \times {\mathbb R}^3$ IR. This is consistent with the AdS$_5$ asymptotic analysis, in which density deformation is more relevant compared to the magnetic one.

Finally, we will end this section with numerical solutions that interpolate between the AdS$_2$ or the AdS$_3$--IR and the AdS$_5$--UV. The interested reader will find relevant details in appendix B, explaining how we construct the numerical solutions. Here we will just present a few numerical results demonstrating our claim. 
\begin{figure}[ht!]
\begin{center}
{\includegraphics[width=0.6\textwidth]{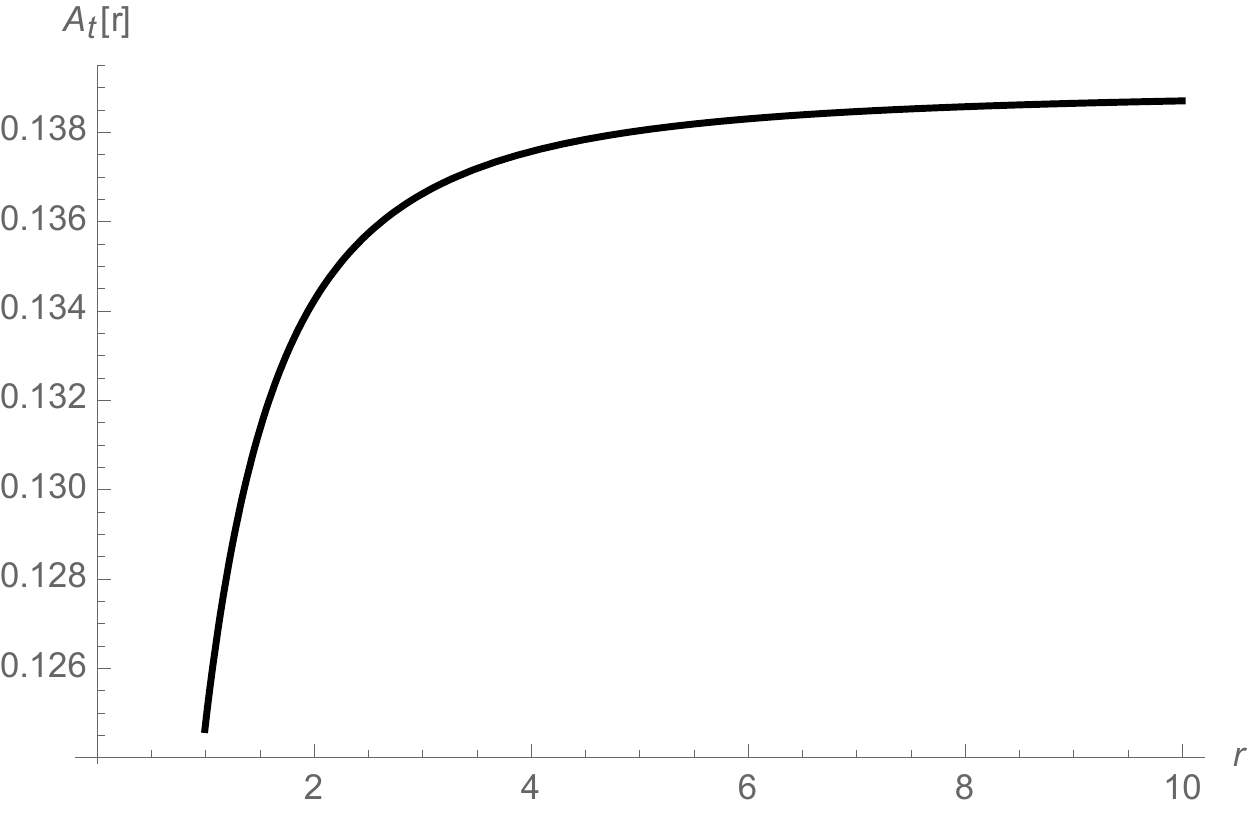}}
{\includegraphics[width=7.4cm]{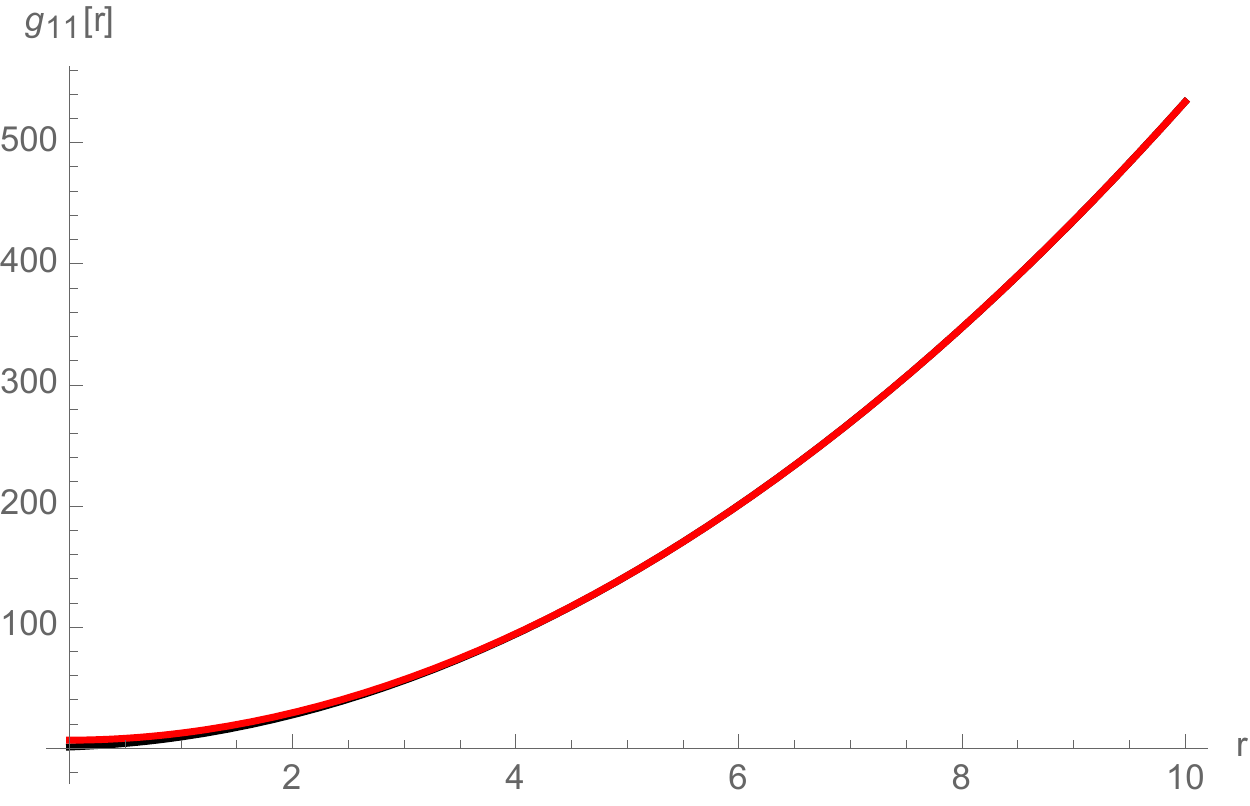}}
{\includegraphics[width=7.4cm]{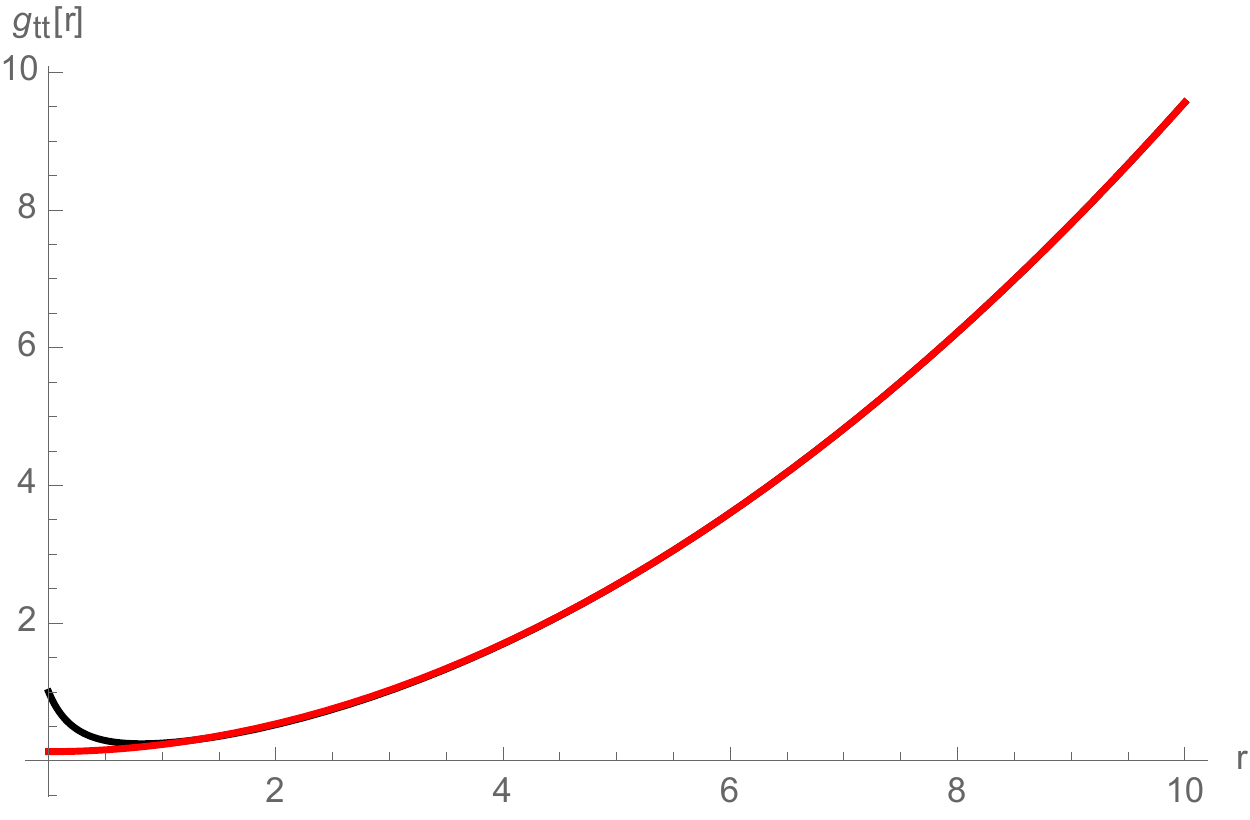}}
\caption{\small The numerical interpolating solution is shown here. We have obtained this particular solution with the following values: $C_1 =1 $, $C_H =1$ and $Q_e=0.5$ (see (\ref{expn2}), (\ref{pertsol1})) and (\ref{bcksol1t}), in units of the AdS$_2$ radius. The $C_H$ mode is irrelevant, see (\ref{bcksol1t}). The numerical integration is performed from $r = 10^{-3}$ to $r = 10$.} \label{AdS2interpolate}
\end{center}
\end{figure}
\begin{figure}[ht!]
\begin{center}
{\includegraphics[width=0.6\textwidth]{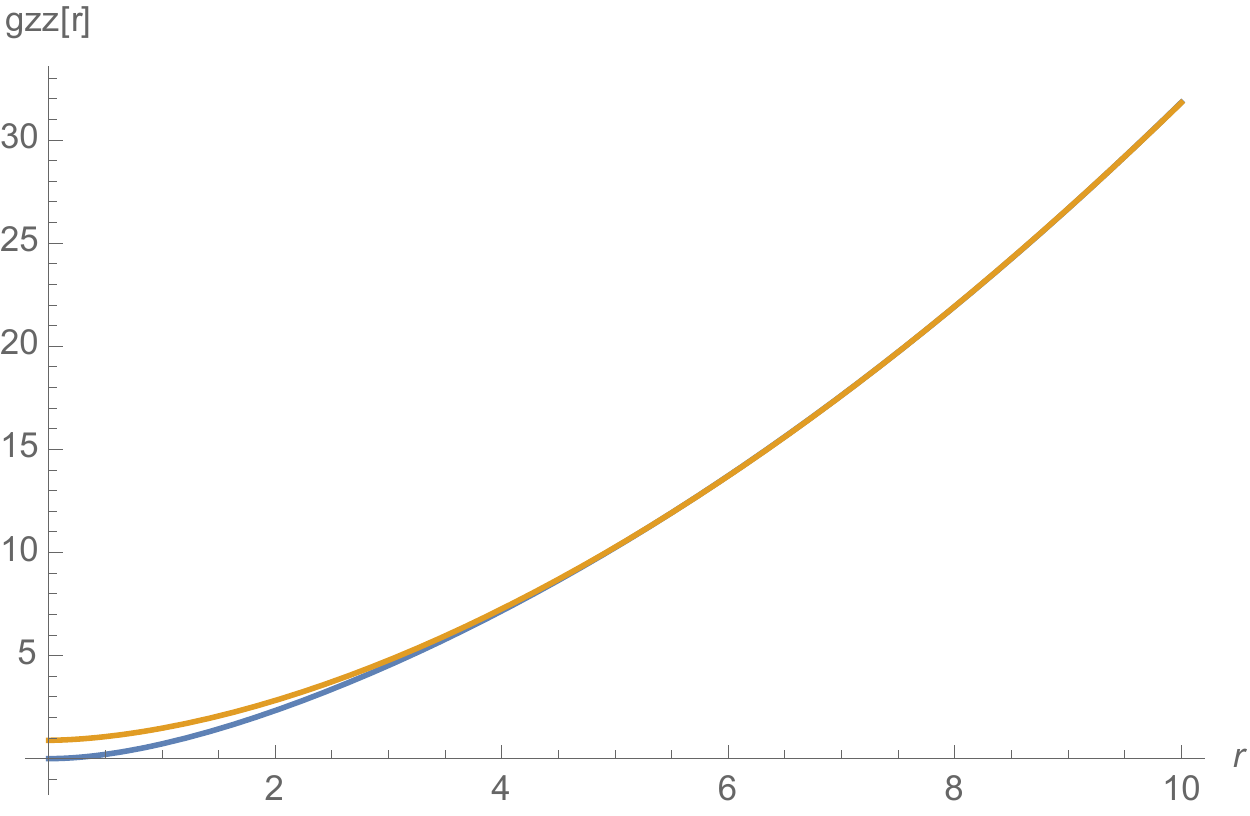}}
{\includegraphics[width=7.4cm]{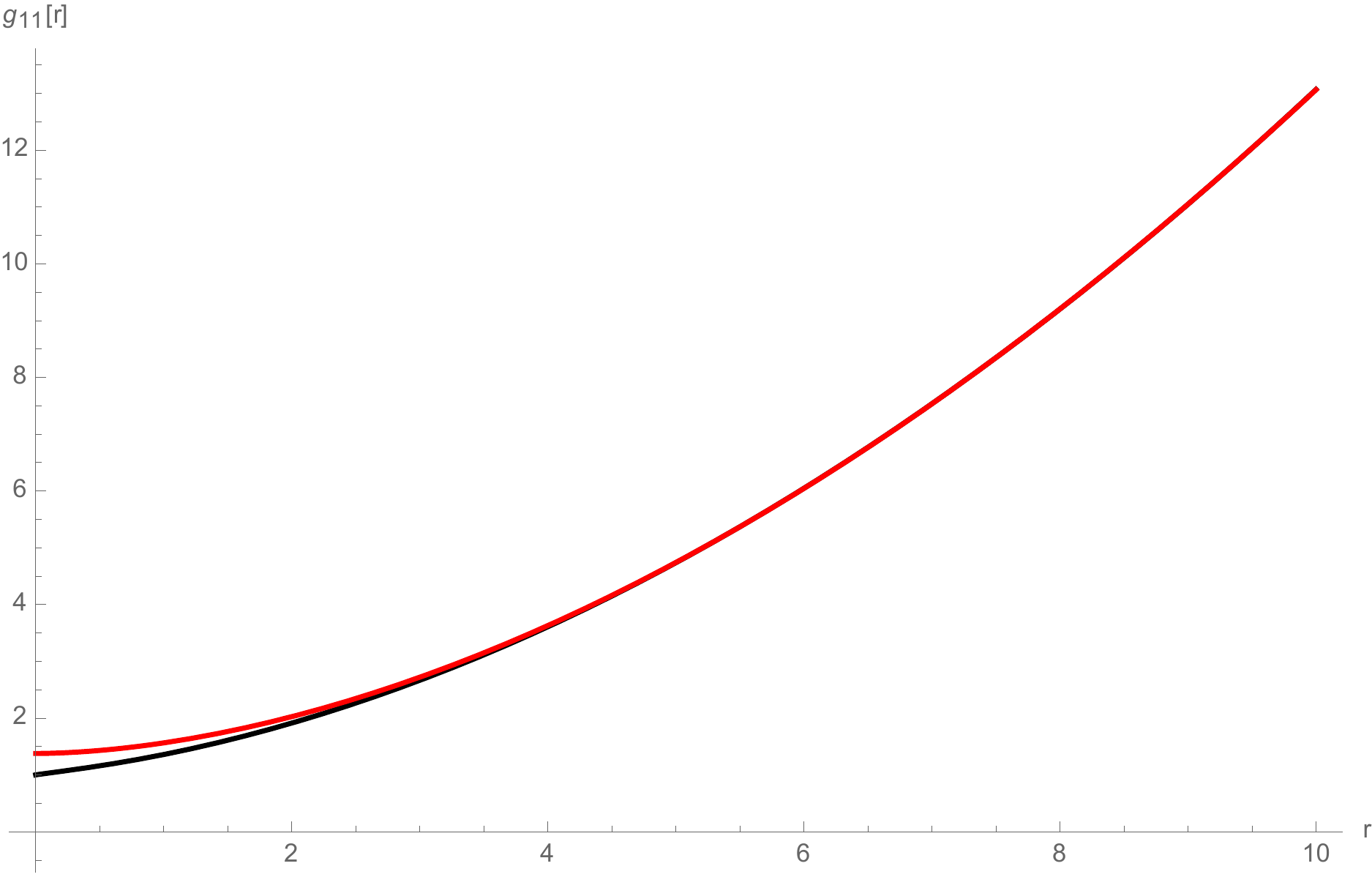}}
{\includegraphics[width=7.4cm]{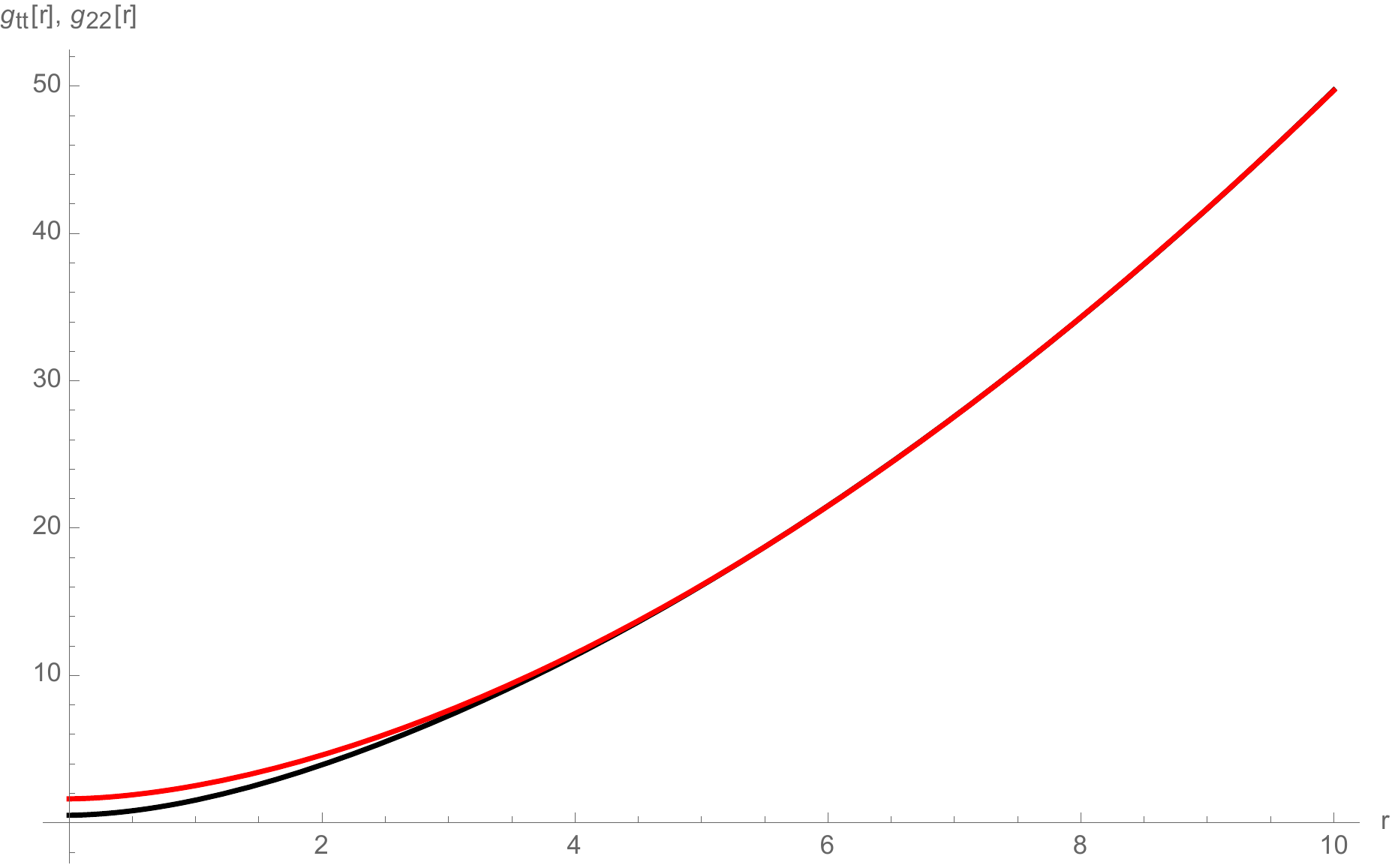}}
\caption{\small The numerical interpolating solution is shown here. We have obtained this particular solution with the following values: $D_1 =-2 $, $C_H =0.5$ and $Q_e=0.8$ (see (\ref{bcksol2nd1}), (\ref{bcksol2mag})) and (\ref{magflucs}), in units of the AdS$_3$ radius. The $C_H$ mode is irrelevant, see (\ref{bcksol2mag}). The numerical integration is performed from $r = 10^{-3}$ to $r = 10$.} \label{AdS3interpolate}
\end{center}
\end{figure}

First, let us consider the AdS$_2\times {\mathbb R}^3$ to AdS$_5$ flow. We have outlined the details, containing admissible boundary conditions, in equations (\ref{expn2}), (\ref{pertsol1})) and (\ref{bcksol1t}). As a representative example, we choose $C_1 =1 $, $C_H =1$ and $Q_e=0.5$, all in units of the AdS$_2$-radius. The corresponding numerical solution\footnote{It was pointed out to us by Javier Tarr\'{i}o that the extremal case can be analytically solved and the solutions are given in \cite{Pal:2012zn, Tarrio:2013tta}.} is shown in figure \ref{AdS2interpolate}. The other interpolating solution from AdS$_3\times {\mathbb R}^2$ to AdS$_5$ is shown in figure \ref{AdS3interpolate}. Here also, we have chosen a representative example, in which $D_1 =-2 $, $C_H =0.5$ and $Q_e=0.8$, in units of the AdS$_3$ radius.

\subsection{Bianchi from DBI: Electric Field}

In this section we will present a particular anisotropic solution that falls under the Bianchi type-V class. The solution is equivalent to an ${\rm AdS}_2 \times {\rm EAdS}_3$ geometry. As before, the general metric ansatz that we will assume is of the following kind
\be 
\begin{split}
& ds^2 = -g_{tt}(r) dt^2 + g_{rr}(r) dr^2 + g_{xx}(r) dx^2 + g_{yy}(r,x) dy^2+  g_{zz}(r,x) dz^2 \ , \\
& A_{\mu} = \{A_t(r),0,0,0,0 \},~~\text{with}~~ A_t(r) = Q_e  r^{\alpha_1} \ . 
\end{split}
\ee
Here we are allowing for the possibility that either one or both of the metric coefficients $g_{yy},~g_{zz}$ are considered to be functions of the coordinates $r,~x$, where the other ones are only functions of the radial coordinate.

For this Bianchi type-V solution  the specific metric ansatz further takes the form:
\be 
\begin{split}
& g_{tt}(r)= L_1 r^{\alpha} \ , ~ g_{rr} (r)= L_1 r^{\beta} \ ,~ g_{xx }(r) = L_2 r^{\delta} \ , \\
& g_{yy} (r,x)=  L_2 r^{\delta} e^{-2x} \ ,~ g_{zz}(r,x) =  L_2 r^{\delta} e^{-2x} \ .
\end{split}
\ee
The algebra of the the Bianchi type-V, generated by the generators of the $3$-dimensional subspace spanned by the coordinates $x,y,z$ is:
\be 
\begin{split}
& \zeta_1 = \partial_{x}+y \partial_{y}+  z\partial_{z} \ , \quad \zeta_2 =  \partial_{y} \ , \quad \zeta_3 =  \partial_{z} \ , \\
& [\zeta_1,\zeta_2] = -\zeta_2 \ , \quad [\zeta_2,\zeta_3] = 0 \ , \quad [\zeta_3,\zeta_1]= \zeta_3 \ .
\end{split}
\ee
The corresponding solution is given by
\be
\begin{split}
& \alpha_1 = 1 \ ,~~ \delta =0 ,~~\beta = -2 \ ,~~\alpha=2 \ , \\
&\Lambda =  \frac{2 L_1^2-L_1 -3 Q_e^2}{L_2 Q_e^2} \ ,~~\tau = \frac{(L_2-2 L_1) \sqrt{(L_1-Q_e) (L_1+Q_e)}}{\kappa^2 L_2 Q_e^2} \ .
\end{split}
\ee
The metric takes the form:
\be 
ds^2 = L_1 \left(-r^2 dt^2 + {dr^2 \over r^2} \right) + L_2 \left( dx^2 + e^{-2x} dy^2+ e^{-2x} dz^2\right) \ ,
\ee
which, after the following coordinate transformation
\be 
x = \log u \ ,
\ee
looks like:
\be 
ds^2 = L_1 \left(-r^2 dt^2 + {dr^2 \over r^2} \right) + L_2 \left( {du^2 + dy^2 +dz^2 \over u^2}\right) \ .
\ee
Here we can explicitly see that the $\{t,~r \}$ corresponds to an ${\rm AdS}_2$, while $\{u,~y,~z \}$ represents an ${\rm EAdS}_3$. As before, in the limit $L_2=2 L_1$, the DBI sector decouples and we obtain the same  ${\rm AdS}_2 \times {\rm EAdS}_3$ solution sourced entirely by a negative cosmological constant. Note that, for arbitrary $d$ we similarly obtain an AdS$_2 \times {\rm EAdS}_{d-1}$ solution.

Within the same ansatz, there is another algebraic solution, described by:
\be 
\beta=\alpha-2 \ ,~\delta=0 \ ,~\alpha_1=\alpha \ ,~~ \Lambda = \frac{\frac{2L_1^2}{\alpha ^2 Q_e^2}-3}{L_2} \ ,~~\tau = -2L_1\frac{\sqrt{L_1^2-\alpha ^2 Q_e^2}}{\alpha ^2 \kappa ^2 L_2 Q_e^2} \  .
\ee
The corresponding line-element, written in $ x = \log u $, $r = e^v$ plane, takes the form:
\be 
ds^2 = L_1 e^{\alpha v}\left(- dt^2 + dv^2 \right) + L_2 \left( {du^2 + dy^2 +dz^2 \over u^2}\right) \ .
\ee
Upon the following further coordinate transformation, and an analytic continuation, all given by
\begin{eqnarray}
&& \tilde{v} = \frac{1}{\alpha} \left[ e^{\frac{\alpha}{2} \left(v+t \right)} + e^{\frac{\alpha}{2} \left(v - t \right)} \right] \ , \quad \bar{t} = \frac{1}{\alpha} \left[ e^{\frac{\alpha}{2} \left(v+t \right)} - e^{\frac{\alpha}{2} \left(v - t \right)} \right] \ , \\
&& \bar{t} = i \tilde{t} \ , \quad z = i \tilde{z} \ , \quad {\rm and} \quad \tilde{Q}_e = i Q_e \ ,
\end{eqnarray}
the line-element turns out to be:
\begin{eqnarray}
ds^2 = L_1 \left( d\tilde{t}^2 + d \tilde{v}^2 \right) + L_2 \left(\frac{du^2 + dy^2 - d\tilde{z}^2}{u^2} \right) \ .	 
\end{eqnarray}
Thus, we get the known AdS$_3\times\R^2$ solution, already given in (\ref{adsqmd}).

\subsection{Anisotropy with Magnetic Field: ${\rm AdS}_3$ Solution}

We will now discuss anisotropic solution sourced by magnetic field. Let us begin with the ansatz:
\be 
\begin{split}
& ds^2 = -g_{tt}(r) dt^2 + g_{rr}(r) dr^2 + g_{xx}(r) dx^2 + g_{yy}(r) dy^2+  g_{zz}(r,y) dz^2 \ , \\
& A_{\mu} = \{0,0,0,0,A_z(y) \} \ ,~~\text{with}~~ A_z(y) = Q_m  e^{-\alpha_1 y} \ ,
\end{split}
\ee
with
\be 
\begin{split}
& g_{tt}(r)= L_1 r^{\alpha}  \ , ~ g_{rr} (r)= L_1 r^{\beta} \ , ~ g_{xx }(r) = L_1 r^{\alpha} \ , \\ 
&g_{yy} (r)=  L_2 r^{\delta} \ ,~ g_{zz}(r,y) =  L_2 r^{\delta} e^{-2y} \ .
\end{split}
\ee
The solution is described by
\be
\begin{split}
&\alpha=-\beta=2 \ , ~ \alpha_1=1 \ ,~ \delta=0 \ , \\
&\Lambda = \frac{L_2 (L_1-2 L_2)-3 Q_m^2}{L_1 Q_m^2} \ ,~\tau = \frac{(2 L_2-L_1) \sqrt{L_2^2+Q_m^2}}{\kappa ^2 L_1 Q_m^2} \ .
\end{split}
\ee
With a variable change of $y = \log u$, the corresponding metric can be written as:
\be 
ds^2 = L_1 \left(-r^2 dt^2 + {dr^2 \over r^2}+ r^2 dx^2 \right) + L_2 \left( {du^2 +dz^2 \over u^2}\right) \ .
\ee
Here we can explicitly see that the $\{t,~r,~x \}$ part describes an ${\rm AdS}_3$, whereas $\{u,~z \}$ part describes an ${\rm EAdS}_2$. Setting $L_1=2 L_2$ again decouples the DBI-matter. In general $(d+1)$-bulk dimensions, one obtains an ${\rm AdS}_{d-1} \times {\rm EAdS}_2$ solution.

The other algebraic solution, which is given by
\be
\begin{split}
& \alpha=0 \ , ~ \alpha_1=1 \ ,~ \delta=0 \ , \\
&  \Lambda = \frac{L_2}{Q_m^2} \ ,~~\tau =-\frac{\sqrt{L_2^2+Q_m^2}}{\kappa ^2 Q_m^2} \ ,
\end{split}
\ee
with the corresponding line-element:
\be 
\begin{split}
ds^2 = & L_1 \left(- dt^2 + r^{\beta} dr^2  + dx^2\right) + L_2 \left( dy^2+ e^{-2y} dz^2\right)
\\  = & L_1 \left(- dt^2 +dv^2  + dx^2\right) + L_2 \left( {du^2+ dz^2\over u^2}\right) \ . 
\end{split}
\ee
%

\section{Partially Filling Branes}

In the spirit of emulating explicit D-brane sources, we will briefly comment on the case when the D-brane is partially filling {\it e.g.}~an AdS$_5$ spacetime. To simplify the problem, we will further {\it smear} the partially-filling branes along the transverse directions, such that the resulting Einstein equations still remain ordinary differential equations. We intend to study the following action:
\begin{eqnarray}
&& S_{\rm full} = S_{\rm gravity} + S_{\rm DBI} \ , \label{act1p} \\
&& S_{\rm gravity} = \frac{1}{2 \kappa^2} \int d^{d+1} x \sqrt{- {\rm det} g_{(d+1)}} \ \left(R - 2 \Lambda \right) \ , \label{act2p} \\
&& S_{\rm DBI} = - \tau \int d^{p+1} x \sqrt{- {\rm det} \left( g_{(p+1)} + F \right) } \int d^{d-p} x \ . \label{act3p}
\end{eqnarray}
Note that, in writing the matter action, we have manifestly lost covariance in the $(d-p)$-directions, those directions are, however, still symmetries of the system. The field $F$ is a $U(1)$-gauge field living on the $(p+1)$-dimensional brane. An AdS$_{d+1}$-solution is obtained if $\Lambda = - d (d-1)/ 2 L^2$, with $\tau = 0 $, where $L$ represents the radius of AdS. In the above, $g_{(p+1)}$ is essentially the components of $g_{(d+1)}$, restricted on to the worldvolume directions of the brane. We are certainly assuming that the brane embedding profile is trivial.

The Maxwell equation remains same as in (\ref{max}). The Einstein equations of motion split into two parts:
\begin{eqnarray}
\left. R_{\mu\nu} - \frac{1}{2} \left( R - 2 \Lambda \right) g_{\mu\nu} \right|_{(p+1)} =  T_{\mu\nu}  \ ,  \quad \left. R_{\mu\nu} - \frac{1}{2} \left( R - 2 \Lambda \right) g_{\mu\nu} \right|_{(d-p)} =  0 \ ,\label{einp}
\end{eqnarray}
where, as before, 
\begin{eqnarray}
&& T^{\mu\nu}  = - \left( \kappa^2\tau \right) \frac{\sqrt{- {\rm det} \left( g_{(p+1)} + F \right)}}{\sqrt{ - {\rm det} g_{(d+1)}}} \S^{\mu\nu} \ , \label{Tdbip}\\
&& \S^{\mu\nu} = \left( \frac{1}{g_{(p+1)} + F} \cdot g_{(p+1)} \cdot \frac{1}{g_{(p+1)} - F} \right)^{\mu\nu} \label{symp}
\end{eqnarray}
For example, one finds the following solutions:
\begin{eqnarray}
&& d=4\ , \ p=3 \ , \quad ds^2 = L_1 \left( - r^2 dt^2 + \frac{dr^2}{r^2} + r^2 \left( dx_1^2 + dx_2^2 \right)\right) + dx_3^2  \ , \\
&& d=4\ , \ p=2 \ , \quad ds^2 = L_1 \left( - r^2 dt^2 + \frac{dr^2}{r^2} + r^2 dx_1^2 \right) + dx_2^2 + dx_3^2  \ , \\
&& d=4\ , \ p=1 \ , \quad ds^2 = L_1 \left( - r^2 dt^2 + \frac{dr^2}{r^2} \right) + dx_1^2 + dx_2^2 + dx_3^2  \ .
\end{eqnarray}
These are AdS$_4\times {\mathbb R}$, AdS$_3\times {\mathbb R}^2$ and AdS$_2\times {\mathbb R}^3$, respectively. The transverse directions to the brane source decouples and becomes an ${\mathbb R}^{d-p}$. Interestingly, within the scaling ansatz, these are the only solutions. Furthermore, the solutions are also non-perturbative in back-reaction (preserving the AdS-asymptotics), which is best reflected in how the cosmological constant and the radius of curvature are related to the other parameters in the theory:
\begin{eqnarray}
\Lambda = - \frac{\left( p+1 \right) }{2} \kappa^2 \tau \ , \quad L_1 = \frac{p}{\kappa^2\tau} \ .
\end{eqnarray}
Clearly, the formula for $L_1$ does not have a well-defined $\tau \to 0$ limit. Unfortunately, in this case, there is no exact solution within a scaling ansatz once the gauge fields on the DBI-worldvolume are turned on. We note, however, that akin to the ``ABJM-case" studied in \cite{Faedo:2014ana}, at non-vanishing density, the IR may asymptote to an AdS$_2$ in a suitable radial expansion. We leave this for future exploration.

\section{Acknowledgements}

AK would like to thank A.~F.~Faedo, D.~Mateos, C.~Pantelidou and J.~Tarr\'{i}o for the wonderful collaborations in \cite{Faedo:2014ana, Faedo:2015urf} and for numerous illuminating discussions. We specially thank Javier Tarr\'{i}o for many useful comments on the first version of this article. We also thank Sandipan Kundu for discussions on recent developments in AdS$_2$-holography. NK is supported by JSPS Grant-in-Aid for Scientific Research (A) No.16H02182. He further acknowledges the support received from his previous affiliation HRI, Allahabad, during the course of this work and the hospitality of SINP, Kolkata during a visit that initiated this work. We sincerely acknowledge the generous and unconditional support of the People of India towards research in basic sciences.

\appendix

\section{Perturbation Around AdS: Various Cases}

In this appendix, we will collect some useful results and elaborate on the various modes that appear as pure gravity fluctuations ({\it i.e.}~without any sources). Thus we consider fluctuations in metric components only, and solve Einstein equations. From this, one is able to extract {\it e.g.}~the stress energy tensor of the dual field theory. We will review this exercise in three distinct cases: the UV AdS$_5$, the magnetically driven AdS$_3 \times \mathbb R^2$ and the density driven AdS$_2 \times \mathbb R^3$.

Let us begin with the UV AdS$_5$ case. In the absence of any DBI-source (the fundamental matter), the solution is characterized by a negative cosmological constant: $\Lambda = - 6 / L_1$. Within the same truncation, {\it i.e.}~keeping $\tau=0$, we can consider linear fluctuations and solve Einstein equations to obtain:
\begin{eqnarray}
&& g_{tt} = r^2 L_1 \left(1 + \delta g_{tt} \right)  \ , \quad g_{xx} = r^2 L_1 \left(1 + \delta g_{xx} \right)  \ , \quad g_{rr} = r^{-2} L_1 \left(1 + \delta g_{rr} \right)  \ , \label{deltag1} \\
&&  \delta g_{tt} = \frac{\varepsilon}{r^4} \ , \quad \delta g_{xx} = \frac{p}{r^4} \ , \quad \delta g_{rr} = 0 \ , \quad {\rm with} \quad \varepsilon + 3 p = 0  \ . \label{deltag2}
\end{eqnarray}
The last relation is the rather familiar equation of state for a $(3+1)$-dimensional CFT, in which $\varepsilon$ and $p$ correspond to the energy and pressure, respectively. Also, setting $\delta g_{rr} = 0 $ is a gauge choice. The linearized Einstein equations do not have any other non-trivial solution.\footnote{It can also be checked trivially, that, in the presence of a non-vanishing $\tau$, with no other fields turned on, the physics is identical. It only changes the cosmological constant which is now given by $\Lambda = - \frac{6}{L_1} - \kappa^2 \tau $.}

Let us now discuss the magnetically driven AdS$_3 \times \mathbb R^2$ case. The solution, already described in (\ref{adsqmd}), is characterized by the following cosmological constant, and DBI-tension:
\begin{eqnarray}
\Lambda = - \frac{1}{L_1}\left( \frac{2 L_2^2}{Q_m^2} + 3\right) \ , \quad \tau = \frac{2 L_2 \sqrt{Q_m^2 + L_2^2}}{ L_1 \kappa ^2 Q_m^2} \ .
\end{eqnarray}
Linearizing and solving Einstein equations now yields:
\begin{eqnarray}
&& g_{tt} = r^2 L_1 \left(1 + \delta g_{tt} \right)  \ , \quad g_{xx} = r^2 L_1 \left(1 + \delta g_{xx} \right)  \ , \quad g_{rr} = r^{-2} L_1 \left(1 + \delta g_{rr} \right)  \ , \\
&& g_{yy} = L_2 \left(1 + \delta g_{yy} \right) \ , \\
&&  \delta g_{tt} = \frac{\alpha_t}{r^\Delta} \ , \quad \delta g_{xx} = \frac{\alpha_x}{r^\Delta} \ , \quad \delta g_{yy} = \frac{\alpha_y}{r^\Delta} \ , \quad \delta g_{rr} = 0  \ ,
\end{eqnarray}
with
\begin{eqnarray}
\Delta = 2 \ , \quad \alpha_t = \varepsilon \ , \quad \alpha_x = p_x \ , \quad \alpha_y = 0 \ , \quad {\rm with} \quad \varepsilon + p_x =  0 \ .
\end{eqnarray}
In the above, $\varepsilon$, $p_x$ and $p_y$ are energy, pressure parallel and perpendicular to the magnetic field, respectively. The equation of state is also reminiscent of an $(1+1)$-dimensional CFT.

There are also other modes, which we write down for completeness (working in the $\delta g_{rr} = 0$ choice):
\begin{eqnarray}
&& \Delta = 1 \pm \frac{\sqrt{\frac{19}{3} + \frac{34}{3} Q_m^2 + 5 Q_m^4}}{1 + Q_m^2} \ , \\
&& \frac{\alpha_t}{\alpha_y}  = \mp  \frac{1}{8 + 6 Q_m^2} \left[  \sqrt{57 + 102 Q_m^2 + 45 Q_m^4} \pm \left( 13 + 9 Q_m^2\right) \right] = \frac{\alpha_x}{\alpha_y} \ .
\end{eqnarray}
Interestingly, $\Delta_{\pm}$ corresponds to a relevant and an irrelevant mode with reference to the AdS$_3$ conformal fixed point.

A similar exercise can be carried out at the AdS$_2\times \mathbb R^3$ fixed point, which is described by(\ref{adsqe}):
\begin{eqnarray}
\Lambda = - \frac{L_1}{Q_e^2} \ , \quad \tau = \frac{\sqrt{L_1^2 - Q_e^2 }}{Q_e^2 \kappa^2} \ .
\end{eqnarray}
Linearizing and solving Einstein equations now yields:
\begin{eqnarray}
&& g_{tt} = r^2 L_1 \left(1 + \delta g_{tt} \right)  \ , \quad g_{xx} = r^2 L_1 \left(1 + \delta g_{xx} \right)  \ , \quad g_{rr} = r^{-2} L_1 \left(1 + \delta g_{rr} \right)  \ , \\
&& g_{yy} = L_2 \left(1 + \delta g_{yy} \right) \ , \\
&&  \delta g_{tt} = \frac{\alpha_t}{r^\Delta} \ , \quad \delta g_{xx} = \frac{\alpha_x}{r^\Delta} \ , \quad \delta g_{yy} = \frac{\alpha_y}{r^\Delta} \ , \quad \delta g_{rr}  = \frac{\alpha_r}{r^\Delta}  \ ,
\end{eqnarray}
where the various modes are:
\begin{eqnarray}
&& \Delta = 1 \ , \quad \alpha_y + \frac{\alpha_x}{2} = 0 \ , \quad \alpha_t + \alpha_r = 0 \ , \label{ads2fluc1} \\
&& \Delta = 2 \ , \quad \alpha_y = 0 = \alpha_x  \ , \quad \alpha_t + \alpha_r = 0 \ , \label{ads2fluc2} \\
&& \Delta = -1 \ , \quad \alpha_x = \alpha_y \ , \quad \frac{\alpha_t}{\alpha_y} = \frac{1}{3} - \frac{Q_e^2}{L_1^2} \ ,  \quad \frac{\alpha_r}{\alpha_y} = \frac{8}{3} - \frac{2 Q_e^2}{L_1^2} \ . \label{ads2fluc3}
\end{eqnarray}
Note that, there are a couple of relevant modes and an irrelevant one, as viewed from the AdS$_2$--fixed point. Also note that, in choosing $\delta g_{rr} = 0$, one would have missed the irrelevant mode altogether. This is unlike the other two cases discussed above, {\it i.e.}~setting $\delta g_{rr} = 0$ does not loose any information for those.

\section{Constructing Interpolating Solutions: Numerical}

In this appendix we consider constructing numerical interpolating solutions between the various fixed points discussed in section \ref{sect:sold4}. Our goal is to demonstrate that the deep IR solution is indeed AdS$_2 \times {\mathbb R}^3$ (electric), or AdS$_3 \times {\mathbb R}^2$ (magnetic). We show this by numerically integrating, using Mathematica, outwards from the near horizon AdS$_2 \times {\mathbb R}^3$ (or AdS$_3 \times {\mathbb R}^2$) IR and establishing that the system asymptotes to an AdS$_5$--UV.

\subsection{AdS$_2 \times {\mathbb R}^3$ to AdS$_5$}

The metric corresponding the interpolating geometry is of the form:
\be 
ds^2=  L_1 \left(-g_{tt}(r) dt^2 + {dr^2 \over g_{tt}(r)}\right) + g_{11}(r) (dx^2 + dy^2+dz^2) \ .
\ee
Here $r$ is the radial coordinate and $r \rightarrow 0, \infty$ corresponds to the IR and the UV, respectively. The IR is of the AdS$_2 \times {\mathbb R}^3$ type, specified by 
\begin{eqnarray} 
&& g_{tt}(r) =r^2 \ ,  ~~ g_{11}(r) = 1 \ . \label{bcksol1} \\
&& A_t(r) = Q_e~ r \ . \label{bcksol2}
\end{eqnarray}
For completeness, let us also recall that this solution is further characterized by:
\be 
\Lambda =  -\frac{L_1}{Q_e^2} \ , ~~ \tau = \frac{\sqrt{L_1^2-Q_e^2}}{\kappa ^2 Q_e^2} \ .
\ee

To proceed further, we will choose the unit $L_1 =1$. The IR is now an one-parameter family of solutions, characterized by $Q_e$. Now, starting with the AdS$_2 \times {\mathbb R}^3$ IR in eq.\eqref{bcksol1}, eq.\eqref{bcksol2} we show that, by adding a suitable perturbation which grows in the UV, this solution is matched to an AdS$_5$--UV. The perturbation is given by 
\be \label{expn1}
\begin{split}
g_{tt} (r) &= r^2 \left(1 + \epsilon~ \delta g_{tt} (r) \right) \\
g_{11} (r)  &= 1 + \epsilon~ \delta g_{11} (r)\\
A_t (r) &= Q_e~r\left(1 + \epsilon~ \delta A_{t} (r) \right)
\end{split}
\ee
with 
\be \label{expn2}
\begin{split}
\delta g_{tt} (r)= C_1~ r^{\nu} \ , ~~
\delta g_{11} (r)=C_2~ r^{\nu} \ , ~~
\delta A_{t} (r)=C_3~ r^{\nu} \ .
\end{split}
\ee
where $C_1,C_2,C_3$ are constants to be determined. Note that the expansion in eq.\eqref{expn1} is a perturbation in $r^{\nu}$, and we have kept a book-keeping parameter $\epsilon$ to determine the order in that expansion, and later we will set this parameter to be unity. Substituting eq.\eqref{expn1}, \eqref{expn2} back in the equations of motion and solving them upto linear order in $\epsilon$ allows us to obtain a perturbation that grows towards UV, which is given by
\be \label{pertsol1}
\nu = 1 \ ,~~C_3 = C_1 \left(\frac{6}{3 Q_e^2-7}+\frac{3}{2}\right) \ ,~~C_2 = \frac{6 C_1}{3 Q_e^2-7} \ .
\ee
As we can see from the above expression, we have a free tunable parameter $C_1$ which we will have to ultimately fix for the numerical interpolation.

Before proceeding further, it should be noted that the perturbations obtained in eq.\eqref{expn1}, \eqref{expn2}, \eqref{pertsol1} are for an extremal ({\it i.e.}~zero temperature) near horizon geometry of the form AdS$_2 \times {\mathbb R}^3$. A near extremal solution (i.e. small but finite temperature) eq.\eqref{bcksol1} will be characterized by
\be \label{bcksol1t}
g_{tt}(r) =r^2 \left(1+ {C_H \over r^2} \right) \ ,  ~~ g_{11}(r) = 1 \ .
\ee
The free parameter $C_H$ sets the Hawking temperature. However, temperature deformation is irrelevant towards the UV, and it will die down as we move out from $r \sim 0$ towards larger $r$. Therefore, at least for small enough temperature ($C_H < 1 $), the same perturbation, as in eq.\eqref{expn1}, \eqref{expn2}, \eqref{pertsol1}, will be strong enough to drive the near horizon and near extremal electric solution to as asymptotic AdS$_5$. Finally, for numerically integrating out the set of second order differential equations, starting from deep IR, one must provide two initial conditions for each of the three variables: $g_{tt},~ g_{11}, A_t$. These are provided in accord with the forms as written in eq.\eqref{expn1}, \eqref{expn2}, and \eqref{pertsol1}.

\subsection{From AdS$_3 \times {\mathbb R}^2$ to AdS$_5$}

A very similar analysis can be done for the AdS$_3 \times {\mathbb R}^2$ to AdS$_5$ interpolation. The general metric is of the form:
\be 
ds^2=  L_1 \left(-g_{tt}(r) dt^2 + {dr^2 \over g_{tt}(r)} +g_{22}(r) dz^2\right) + L_2 g_{11}(r) (dx^2 + dy^2) \ . 
\ee
The AdS$_3 \times {\mathbb R}^2$--IR is given by
\be \label{bcksol2nd1}
g_{tt}(r) = r^2 \left(1+ {C_H \over r^2}\right) \ ,  ~~ g_{11}(r) = 1 \ ,  ~~g_{22}(r) =r^2 \ ,  ~~A_x(y) = Q_m~ y \ ,
\ee
with 
\be 
\Lambda = \frac{-\frac{2 L_2^2}{Q_m^2}-3}{L_1} \ ,~~\tau = \frac{2 L_2 \sqrt{L_2^2+Q_m^2}}{\kappa ^2 L_1 Q_m^2} \ .
\ee
We will work in units where $L_1=1,L_2=2$.

The corresponding perturbation that grows towards UV is of the following form:
\be \label{bcksol2mag}
\begin{split}
& g_{tt} (r) = r^2\left(1+ {C_H \over r^2}\right) \left(1 + D_1~ r^{\nu} \right) \ , ~~
g_{11} (r)  = 1 + D_2~ r^{\nu} \ ,  \\
&g_{22} (r)  = r^2 \left(1 + D_1~ r^{\nu} \right) \ ,
\end{split}
\ee
where
\be \label{magflucs}
\begin{split}
\nu &= \frac{\sqrt{5 Q_m^4+\frac{136 Q_m^2}{3}+\frac{304}{3}}}{Q_m^2+4}-1 \ , \\
D_2&=-\frac{D_1 \left(15 Q_m^2+2 \sqrt{45 Q_m^4+408 Q_m^2+912}+76\right)}{6 Q_m^2+56} \ .
\end{split}
\ee
With these, one can now numerically integrate the set of the differential equations.

\end{document}